\documentclass{article}

\def\Bbb{\bf}

\newcommand{\x}{\underline{x}}
\newcommand{\y}{\underline{y}}

\font\Eul = eufm10

\def\W{\makebox{\Eul{W}}}
\newtheorem{definition}{Definition}
\newtheorem{lemma}{Lemma}
\newtheorem{prop}{Proposition}

\newtheorem{thm}{Theorem}
\newtheorem{remark}{Remark}
\def\build#1_#2^#3{\mathrel{
\mathop{\kern 0pt#1}\limits_{#2}^{#3}}}
\def\W{\makebox{\Eul{W}}}

\font\Eul = eufm7 at 12pt

\def\bC{{\bf C}}
\def\bR{{\bf R}}
\def\bN{{\bf N}}
\def\Im{{\rm Im\,}}
\def\Re{{\rm Re\,}}
\def\Rp{{\bf R}_+}

\def\th{{\rm th\,}}
\def\ch{{\rm ch\,}}
\def\sh{{\rm sh\,}}
\def\AA{{\cal A}}
\def\BB{{\cal B}}
\def\CC{{\cal C}}
\def\DD{{\cal D}}

\def\HH{{\cal H}}

\def\JJ{{\cal J}}

\def\NN{{\cal N}}

\def\RR{{\cal R}}
\def\SS{{\cal S}}
\def\TT{{\cal T}}

\def\VV{{\cal V}}
\def\WW{{\cal W}}

\def\ZZ{{\cal Z}}
\def\wh{\widehat}
\def\wt{\widetilde}
\def\ovl{\overline}
\def \vhi{\varphi}
\def \veps{\varepsilon}
\def\e{{\rm e}}
\def\HB{\hfill\break}

\def\Xdn{X_d^n}
\def\Xdm{X_d^m}
\def\Xcdn{X_d^{(c)n}}
\def\Xcdm{X_d^{(c)m}}
\def\Gd{L_+^\uparrow}
\def\Rep{{\bf \Phi}}

\def\Rpo{(0,\ +\infty)}
\def\Rpc{\Rp}
\def\Rmo{(-\infty,\ 0)}

\def\ifundefined#1{\expandafter\ifx\csname#1\endcsname\relax}

\newcount\apcount
\ifundefined{completionApp} \else \advance \apcount by 1\fi
\ifundefined{BHWextApp} \else \advance \apcount by 1\fi
\ifundefined{GlaserApp} \else \advance \apcount by 1\fi
\def\totapps{\ifcase\apcount \or one \or two \or three \fi}

\title{Analyticity properties and thermal effects for general quantum field
theory on de Sitter space-time.}
\author{{Jacques Bros$^{1}$, Henri Epstein$^{2,3}$ and Ugo Moschella$^{1,3,4}$  }\\
 {$^{1}$Service de Physique Th\'eorique, C.E. Saclay,
91191 Gif-sur-Yvette, France}\\
{$^{2}$C.N.R.S.}\\
{$^{3}$Institut des Hautes Etudes Scientifiques, 91440 Bures-sur-Yvette, 
France}\\
{$^{4}$ Istituto di Scienze Matematiche Fisiche e Chimiche, Via Lucini 3, 22100 Como,}\\
{and INFN sez. di Milano, Italy}}

\begin{document}
\maketitle

\begin{abstract}
We propose a general framework for quantum field theory on the de Sitter 
space-time (i.e. for local field theories whose truncated 
Wightman functions are not required to vanish).
By requiring that the fields satisfy a {\em weak spectral condition}, 
formulated in terms of the analytic continuation properties of their 
Wightman functions,  
we show that a geodesical observer   
will detect in the corresponding  ``vacuum'' a blackbody radiation at 
temperature $T={1}/{2\pi R}$.
We also prove the analogues of the PCT, Reeh-Schlieder and Bisognano-Wichmann
theorems. 
\end{abstract}

\section{Introduction}

It is known that, when  quantizing  fields  on a gravitational background,
it is  generally impossible to characterize the physically relevant 
vacuum states  as the fundamental states for the energy in the usual sense, 
since there is no such thing as a global energy operator. 

In the absence of the analogue of an energy-momentum  spectrum condition 
\cite{[J],[SW]}, several authors have formulated various alternative  
prescriptions to select, among the possible representations (vacua) 
of a quantum field theory, those which can have a meaningful physical 
interpretation;  {\em the adiabatic prescription, the local Hadamard 
condition}, and {\em the conformal criterion},  (see \cite{[BD],[KW]} and 
references therein) have proven to be useful for characterizing linear 
field theories with vanishing truncated $n$-point functions (i.e. 
{\em free field theories}) on various kinds of space-times. 

It is worthwhile to stress immediately that the relevant choice of the 
vacuum of a quantum  field theory  on a curved space-time has striking 
consequences even in the case of free fields: the most celebrated examples 
are the Hawking thermal radiation  on a black-hole background 
\cite{[HA],[HA2],[S]}, the Unruh effect \cite{[U]} and the Gibbons and  
Hawking  thermal effect on a de Sitter space-time \cite{[GH]}.

As regards interacting field theories on a gravitational background,
(i.e. field theories with non-vanishing truncated $n$-point functions),
much less is known. While the property of {\em locality} (or {\em local
commutativity}) of the field 
observable algebra (i.e. the commutativity of any couple of
field observables localized in space-like separated 
regions) remains a reasonable 
postulate for all space-time manifolds which are globally hyperbolic,
the problem of specifying a representation of the field algebra becomes
still more undetermined than in the free-field case.
In the latter, the indeterminacy is confined in the two-point functions
of the fields, namely in the splitting of the given (c-number)
commutators into the correlators at permuted couples of points.
In the general interacting case, 
the indeterminacy of the possible representations is now encoded
(in an unknown way) in the properties of the whole sequence of
$n$-point functions of the fields. If the gravitational background
is only considered as a general (globally hyperbolic and
pseudo-Riemannian) differentiable manifold, this huge indeterminacy
cannot be completely reduced by imposing a (well-justified) 
{\em principle of
stability} which postulates 
the existence at each point of the manifold of a Minkowskian
scaling limit of the theory satisfying the spectral condition
(see \cite{[H]} and references therein); nor can it be 
reduced in an operational way by adding the general  
requirement of a {\em local definiteness criterion} (based on
the principles of local quantum physics \cite{[H]}).
In such a general context, one should however mention the 
recent use of microlocal analysis which has allowed the 
introduction of a {\em wave front set approach to the spectral condition}
\cite{[R],[BFK]}; after having supplied a simple characterization
of the free-field Hadamard states, this promising 
approach has in its program to give information on  
the $n$-point functions of 
interacting fields {\em in perturbation theory}. 

On the other side, starting from the remark that in Minkowskian theories
the spectral condition can be expressed in terms of analyticity
properties of the $n$-point functions in the complexified 
space-time manifold \cite{[J],[SW]}, one can defend the viewpoint
that it may be of particular interest to study quantum field theory
on an {\em analytic} gravitational background. As a matter of fact,
there is one model of analytic curved universe, and actually 
the simplest one, that offers the unique possibility of formulating 
{\em a global spectral condition for interacting fields} which is very
close to the usual spectral condition of Minkowski QFT: this is the 
de Sitter space-time.

The de Sitter space-time can be represented as a $d$-dimensional 
one-sheeted hyperboloid embedded in a Minkowski 
ambient space ${\Bbb R}^{d+1}$ and it can also be seen 
as a one-parameter deformation of a $d$-dimensional Minkowski 
space-time involving a length $R$.
The Lorentz group of the ambient space acts as a relativity group for 
this space-time, and the very existence of this (maximal) 
symmetry group explains the popularity of  the de Sitter universe as 
a convenient simple model for developing techniques of QFT on a 
gravitational background. Moreover, 
there has been a regain of interest in the de Sitter metric   
in the last years, since it has been considered to play a central 
role  in the inflationary
cosmologies (see \cite{[L]} and references therein): 
a possible explanation of phenomena occurring in the very early universe 
then relies on an interplay between space-time curvature and thermodynamics
and a prominent role is played by the mechanisms of symmetry breaking
and restoration in a de Sitter QFT.
  
The geometrical properties of de Sitter space-time and of its 
complexification actually make it possible to formulate a general 
approach to QFT on this universe which closely parallels the Wightman
approach \cite{[J],[SW]} to the Minkowskian QFT. 
In fact, it is not only the existence
of a simple causal structure (inherited from the ambient Minkowski
space) and of a global symmetry group (playing the same role as the
Poincar\'e group) on the real space-time manifold which are similar;
but the complexified manifold itself is equipped with domains which are
closely similar to the tube domains of the complex Minkowski space.
Since these Minkowskian tubes play a crucial role for expressing the spectral
condition in terms of analyticity properties of the $n$-point 
functions of the theory, the previous geometrical remark strongly suggests 
that analogous complex domains might be used for a  
global formulation of the spectral condition in 
de Sitter quantum field theory. 
This approach has been in fact introduced and used successfully
in a study of {\em general two-point functions} on de Sitter space-time
\cite{[Br],[BGM],[BM],[Mo]}.
As a by-product, it has been shown \cite{[BM]} that a satisfactory
characterization of {\em generalized free fields (GFF) on de Sitter
space-time}, including the preferred
family of de Sitter invariant Klein-Gordon field theories (known as
Euclidean \cite{[GH]} or Bunch-Davies \cite{[BuD]} vacua) 
can be given in terms of the
global analytic structure of their two-point functions in the complexified
de Sitter manifold. Moreover, all these theories of GFF were shown to be
equivalently characterized by the existence of thermal properties of
Gibbons-Hawking type, the temperature $T=(2\pi R)^{-1}$ being induced by the
curvature of the space.

In this paper, we will show that the same ideas and methods can be applied
with similar results
to a general approach to the theory of interacting quantum fields in de
Sitter space-time. In fact, we shall work out an axiomatic program 
(already sketched at the end of
\cite{[BM]}) in which the ``spectral condition'' is replaced by 
appropriate {\em  
global analyticity properties of the $n$-point vacuum expectation values of
the fields} (or ``Wightman distributions'') in the complexified de Sitter
manifold. These postulated analyticity properties are similar to those
implied by the usual spectral condition in the Minkowskian case, according
to the standard Wightman axiomatic framework. For simplicity, we shall
refer to them as to the ``weak spectral condition''. 

As a physical support to our weak spectral
condition, we shall establish 
that all interacting fields which belong to this general framework admit a 
Gibbons-Hawking-type
{\em thermal interpretation} with the same specifications as the one
obtained for GFF's in \cite{[BM]}. In spite of this remarkable
interpretative discrepancy with respect to the Minkowskian quantum fields
satisfying the usual spectral condition, we shall see that such basic
structural properties as the PCT and Reeh-Schlieder theorems
are still valid in this general approach to de Sitter QFT. Furthermore, our
global analytic framework also supplies an analytic continuation of the
theory to the ``Euclidean sphere'' of the complexified de Sitter
space-time, which is the analogue of the (purely imaginary time)
``Euclidean subspace'' of the complexified Minkowskian space-time.
We will also show that the Wick powers of generalized free fields 
fit within the framework and we have indication that our approach is relevant
for the study of perturbation theory. 
The latter will be developped elsewhere.

{}From a methodological
viewpoint, one can distinguish (as in the Minkowskian case) two types of
developments which can be called according to a traditional terminology the
``linear'' and ``non-linear programs''.

The {\em linear program}, which deals exclusively with the
exploitation of the postulates of locality, de Sitter covariance and
spectral condition (expressed by linear relations between the various
permuted $n$-point functions, for each fixed value of $n$) results in the
definition of primitive analyticity domains $D_n$ for all the $n$-point
(holomorphic) functions $\makebox {\Eul W}_{n}(z_1,..., z_n)$ of the theory. 
Each domain $D_n$ is an open
connected subset of the topological product of $n$ copies of the complexified de
Sitter hyperboloid. As in the Minkowskian case, each primitive domain $D_n$
is not a ``natural holomorphy domain'', but it turns out that new regions of
analyticity of the functions $\makebox {\Eul W}_{n}$ 
(contained in the respective holomorphy
envelopes of the domains $D_n$ and obtained by geometrical techniques of
analytic completion) yield important consequences for the corresponding
field theories. A specially interesting example is the derivation of
analyticity properties of the functions $ \makebox {\Eul W}_{n}$ 
with respect to any subset of points $ z_i = z_i(t) $ varying
simultaneously on   
complex hyperbolae interpreted as the (complexified) 
trajectories of a given time-like Killing
vector field on the de Sitter universe. The periodicity with respect to the
imaginary part of the corresponding time-parameter $ t $ directly implies the
interpretation of the obtained analyticity properties 
of the functions $ \makebox {\Eul W}_{n} $ 
as a {\em KMS-type condition}; in view of the general
analysis of \cite{[HHW]}, this gives a thermal
interpretation to all the de Sitter field theories considered.
Since the above mentioned analyticity property is completely similar
to the one which emerges 
from the Bisognano-Wichmann results 
in the Minkowskian theory \cite{[BW]} (see our
comments below), we shall call the previous result ``Bisognano-Wichmann
analyticity property of the $n$-point functions''.

The {\em non-linear program}, which exploits 
the Hilbert-space structure of the
theory, relies in an essential way on the (quadratic)
``positivity inequalities'' 
to be satisfied by the whole sequence of $n$-point
Wightman distributions of the fields; these inequalities just express
the existence of the vector-valued distributions defined by the 
action of field operator products on the ``GNS-vacuum state'' of the
theory. An important issue to be recovered 
is the fact that these distributions are
themselves the boundary values of {\em vector-valued holomorphic
functions} from certain complex domains; it is this mathematical fact 
which is directly responsible for such important features of the theory as
the Reeh-Schlieder property. In the Minkowskian case, this vectorial
analyticity is readily obtained from the spectral condition by an
argument based on the Laplace transformation.
Here, we shall apply an alternative method for establishing vectorial
analyticity which directly makes use of the analyticity and positivity
properties of the $n$-point functions. It is based on  
a general study by V.~Glaser \cite{[G1], [G2]} of 
{\em positive-type sequences of holomorphic kernels
in domains of $\bC^m \times \bC^n$},
whereby the analyticity
of the Wightman $n$-point functions ``propagates'' their positivity
properties to the complex domain.
This method is therefore applicable not only to the Minkowskian
and de Sitter QFT but also, in principle,  
to QFT on more general {\em holomorphic (or real-analytic) 
space-time manifolds}
for which the spectral condition would be replaced by an
appropriate (possibly local) version of the analyticity properties of
the Wightman functions.

The structure of the paper is the following: 
in Section 2,  we introduce the notations and  recall 
some properties of the de Sitter spacetime and of its complexification; 
we then formulate our general principles for the 
interacting fields on this universe, giving a special emphasis on the
spectral condition which we propose. 

In Section 3 we explore various consequences of our general
principles which are the analogues of standard results of
the Minkowskian QFT in the Wightman framework. 
In particular, we establish the existence of an analytic 
continuation of the Wightman $n$-point functions to corresponding 
primitive domains of analyticity.
The PCT property is also shown.

In Section 4 we come to the physical interpretation of the spectral condition. 
We first extend the analytic 
aspect of  the  Bisognano-Wichmann theorem \cite{[BW]} 
to the de Sitter case.
Then we show that  the thermal interpretation, already known for
free field theories \cite{[GH],[BM]}, is still valid in this more general case.

In section 5 we prove the validity of the Reeh and Schlieder property.
The proof of the relevant vectorial analyticity is given as an
application of the above mentioned theorem of Glaser.

\ifnum\apcount=0 \else
The paper is ended by \totapps appendices where  we discuss some more 
technical results.\fi

\section{QFT on the de Sitter spacetime: the spectral condition}

We start with some notations and some well-known facts.
The $(d+1)$-dimensional real (resp. complex) Minkowski space is $\bR^{d+1}$
(resp. $\bC^{d+1}$) equipped with the scalar product 
$
x\cdot y =  {x^{(0)}}{y^{(0)}}-{x^{(1)}}{y^{(1)}}-\ldots-{x^{(d)}}y^{(d)}
\label{product}
$ 
with, as usual, $x^2 = x\cdot x$. We thus distinguish a particular
Lorentz frame and denote $e_\mu$ the $(d+1)$-vector with
$e_\mu^{(\nu)} = \delta_{\mu\,\nu}$.
In this special Lorentz frame, we also distinguish the $(e_0,e_d)$-plane 
and the corresponding 
light-like coordinates $u$ and $v$, namely we put:
\begin{equation}
x = (x^{(0)},\vec{\underline{x}},x^{(d)}),\ \ \ \ \  
\vec{\underline{x}} = (x^{(1)},\ldots,x^{(d-1)}),   
\label{xdef}
\end{equation}
\begin{equation}
u = u(x) = x^{(0)} + x^{(d)},\ \ \ \ v = v(x) = x^{(0)} - x^{(d)} ,
\label{udef}
\end{equation}
and we introduce, for each $\lambda = \e^\zeta \in \bC \setminus \{0\}$, the special
Lorentz transformation $[\lambda]$ such that
\begin{equation}
u([\lambda]x) = \lambda u(x),\ \ \ v([\lambda]x) = \lambda^{-1} v(x),\ \ \ \  
\vec{\underline{x}}([\lambda]x) = \vec{\underline{x}}(x), 
\ \ \ \ [\e^\zeta] = \exp \zeta\, e_0\wedge e_d.
\label{speclor}
\end{equation}
The future cone is defined in the real  Minkowski space $\bR^{d+1}$ 
as the subset
$$
V_+ = -V_- = \{x\in \bR^{d+1}\ :\ x^{(0)} > 0,\ \ x\cdot x >0 \}
$$
and the future light cone as $C_+ = \partial V_+ = -C_-$. We denote
$x \le y$ the partial order (called causal order) defined by $\ovl{V_+}$, i.e. 
$x \le y \Leftrightarrow y-x \in \ovl{V_+}$. 
The $d$-dimensional real (resp. complex)
de Sitter space-time with radius $R$ 
is identified with the  subset of the real (resp. complex) 
Minkowski space consisting of the points $x$ such that $x^2 = -R^2$ and is 
denoted $X_{d}(R)$ or simply $X_{d}$ (resp. $X_{d}^{(c)})$. Thus $X_{d}$
is the one-sheeted hyperboloid 
\begin{equation}
X_{d}=X_{d}(R)=
\{x\in {\Bbb R}^{d+1}: {x^{(0)}}^{2}-{x^{(1)}}^{2}-\ldots-{x^{(d)}}^{2}=-R^{2}\}
\label{hyp} 
\end{equation}
The causal order on $\bR^{d+1}$ induces the causal order on $X_d$. 
The future and past shadows of a given event $x$ in $X_{d}$
are given by
$\Gamma^+{(x)}= \{y \in X_{d}: y \geq x \}$, 
$\Gamma^-{(x)}= \{y \in X_{d}: y \le x \}$.
Note that if $x^2 = -R^2$ and $r^2 = 0$,  then $(x+r)^2 = -R^2$ 
is equivalent to $x\cdot r=0$, and remains true if $r$ is replaced
with $t\,r$ for any real $t$ (the same holds in the complex domain.)
Hence the boundary set 
\begin{equation}
{\partial \Gamma}(x) =\{y \in X_{d}: (x-y)^2=0\}
\end{equation}
of $\Gamma^+{(x)}\cup\Gamma^-{(x)}$ is a cone (``light-cone'') with apex $x$,
the union of all linear generators of $X_{d}$ containing the point $x$.
Two events $x$ and $y$ of $X_{d}$ are in ``acausal relation'', or 
``space-like separated'' if $y \not\in \Gamma^+{(x)}\cup\Gamma^-{(x)}$, 
i.e. if $x\cdot y > -R^2$.
The relativity group of the de Sitter space-time, called ``de Sitter group''
in the following, is 
the connected Lorentz group of the ambient Minkowski space, i.e.
$\Gd=SO_{0}(1,d)$ leaving invariant  each of the sheets of the  cone  
$C = C_{+}\cup C_{-}$. The connected complex Lorentz group in $d+1$ dimensions
is denoted $L_+(\bC)$. We denote $\sigma$ the
$\Gd$-invariant volume form on $X_{d}$ given by
\begin{equation}
\int f(x)\,d\sigma(x) = \int f(x)\,\delta(x^2 + R^2)\,
dx^{(0)} \wedge \ldots \wedge dx^{(d)}.
\end{equation}
$\Gd$ acts transitively on $X_d$ and $L_+(\bC)$ on $X_d^{(c)}$.

The familiar forward and backward tubes are defined in 
complex Minkowski space as ${\rm T}_{\pm} = \bR^{d+1} \pm iV_+$, and
we denote 
\begin{equation}
\TT_{+} = \makebox{\rm T}_+\cap X^{(c)}_{d},\;\;\;\;\;
\TT_{-} = \makebox{\rm T}_-\cap X^{(c)}_{d}.
\label{tubi1}
\end{equation}
Since $\overline{{\rm T}_+}\cup
\overline{{\rm T}_-}$ contains the ``Euclidean subspace''
${\cal E}_{d+1}=\{ z=(iy^{(0)},x^{(1)},\ldots, x^{(d)}): (y^{(0)},
x^{(1)},\ldots,  x^{(d)}) \in {{\Bbb R}}^{d+1} \}$ of the complex Minkowski 
space-time ${{\Bbb C}}^{d+1}$, the subset
$\overline{\TT_+}\cup
\overline{\TT_-}$ of $X_d^{(c)}$
contains the  ``Euclidean sphere''
${ S}_{d}=\{z=({iy^{(0)}},{x^{(1)}},\ldots{x^{(d)}}):\;\; 
{y^{(0)}}^{2}+{x^{(1)}}^{2}+\ldots+{x^{(d)}}^{2}=R^{2}\}$.

We denote $\DD(\Xdn)$ (resp. $\SS(\Xdn)$) 
the space of functions on $\Xdn$
which are restrictions to $\Xdn$ of functions belonging to
$\DD(\bR^{n(d+1)})$ (resp. $\SS(\bR^{n(d+1)})$). As in the Minkowskian
case, the Borchers algebra $\BB$ is defined as the tensor algebra
over $\DD(X_d)$. Its elements are terminating sequences 
of test-functions $f = (f_0, f_1(x_1),\ldots,
f_n(x_1,\ldots,x_n),\ \ldots)$, where $f_0 \in {\Bbb C}$
and $f_n \in {\cal D}(X_d^n)$ for all $n \ge 1$,  
the product and $\star$ operations being given by
$$
(fg)_n = 
\sum_{p,\ q \in \bN \atop p+q = n}\ f_p \otimes g_q,\ \ \ \ 
(f^\star)_n (x_1,\ldots,\ x_n) = 
\ovl{f_n (x_n,\ldots,\ x_1)}.
$$
The action of the de Sitter group $\Gd$ on $\BB$ is defined by
$ f \mapsto f_{\{\Lambda_r\}}$, where 
\begin{equation}
f_{\{\Lambda_r\}} = (f_0, f_{1\{\Lambda_r\}},\ldots,f_{n\{\Lambda_r\}},\ldots),\ \ \ \ 
f_{n\{\Lambda_r\}} ({x_{1}},\ldots,x_{n})
=f_{n} ({\Lambda_r}
^{-1}{x_{1}},\ldots,{\Lambda_r}^{-1}x_{n}), 
\label{Lambdaact}
\end{equation}
$\Lambda_r$ denoting any (real) de Sitter transformation.
\vskip 5pt

A quantum field theory (we consider a single scalar field for
simplicity) is specified by a continuous linear functional $\WW$ on
$\BB$, i.e. a sequence $\{\WW_n \in \DD'(\Xdn)\}_{n \in \bN}$ 
where $\WW_0 =1$ and the $\{\WW_n\}_{n >0}$ are 
distributions (Wightman functions) required to possess the following
properties:

\begin{enumerate}
\item (Covariance). Each ${\cal W}_{n}$ is de Sitter invariant, i.e.
\begin{equation}
\langle {\cal W}_{n},\ f_{n\{\Lambda_r\}}\rangle = 
\langle {\cal W}_{n},\ f_{n} \rangle
\label{cov}
\end{equation}
for all de Sitter transformations $\Lambda_r$. (This is equivalent to saying that 
the functional $\cal W$ itself is invariant, i.e. $ {\cal W}(f) = {\cal W}(f_{\{\Lambda_r\}})$
for all $\Lambda_r$).

\item (Locality)
\begin{equation}
{\cal W}_{n}({x_{1}},\ldots,x_{j},x_{j+1},\ldots,x_{n})
={\cal W}_{n}({x_{1}},\ldots,x_{j+1},x_{j},\ldots,x_{n})
\end{equation}
if $(x_{j}-x_{j+1})^{2}<0$.
\item (Positive Definiteness). For each $f \in \BB$, 
$\WW(f^\star f) \ge 0$. Explicitly, 
given $f_{0} \in {{\Bbb C}}, 
f_{1} \in {\cal D}(X_d),\ldots,$
$f_{k} \in {\cal D}(X_d^{k}),$
then
\begin{equation}
\sum_{n,m=0}^{k}\langle
{\cal W}_{n+m},\ f_{n}^\star\otimes f_{m}\rangle\geq 0.
\label{posit}
\end{equation}
\end{enumerate}

As in the Minkowskian case \cite{[W1],[B],[J]}, 
the GNS construction yields a Hilbert space $\HH$, 
a unitary representation $\Lambda_r \mapsto U(\Lambda_r)$ of $\Gd$,
a vacuum vector $\Omega \in \HH$ invariant under $U$, and an operator valued
distribution $\phi$ such that
\begin{equation}
\WW_n(x_1,\ldots,\ x_n) =
(\Omega,\ \phi(x_1)\ldots\phi(x_n)\,\Omega).
\end{equation}
The GNS construction also provides the vector valued distributions
$\Phi_n^{(b)}$ such that  
\begin{equation}
\langle \Phi_n^{(b)},\ f_n \rangle = 
\int f_n(x_1,\ldots,\ x_n)\,\phi(x_1)\ldots\phi(x_n)\,\Omega\,
d\sigma(x_1)\ldots d\sigma(x_n)
\label{vecvaldis}
\end{equation}
and a representation ${  f} \to \Rep({f})$ (by unbounded operators)
of ${\cal B} $ of which the field $\phi$ is a special case:
$ \phi({ f_1})= \int \phi(x)f_1(x) d\sigma(x)
= \Rep\left((0,f_1,0,\ldots)\right)$. For every open set ${\cal O}$ of $X_d$
the corresponding polynomial algebra ${\cal P}({\cal O})$ of the field $\phi$ 
is then defined as the subalgebra of $\Rep({\cal B})$ whose elements 
$\Rep(f_0,f_1,\ldots,f_n,\ldots)$ are such that for all $n\geq 1\;$ 
supp$f_n(x_1,\ldots, x_n)\subset {\cal O}^{n}$.
The set ${\rm D} = {\cal P}(X_d)\Omega$ is a dense subset of ${\cal H}$ and
one has (for all elements $\Rep({f}), \Rep({g}) \in {\cal P}(X_d)$):
\begin{equation}
\WW(f^\star g) = 
(\Rep({f})\Omega,\ \Rep({g})\Omega).
\label{borch}
\end{equation}

The properties 1-3 are literally carried over from the Minkowskian case,
but no literal or unique adaptation exists for the usual spectral
property. In the $(d+1)$-dimensional Minkowskian case, 
the latter is equivalent to the following:
for each $n \ge 2$, $\WW_n$ is the boundary value in the sense of
distributions of a function holomorphic in the tube
\begin{equation}
{\rm T}_n = \{z = (z_1,\ldots,\ z_n) \in \bC^{n(d+1)}\ :\ 
\Im (z_{j+1} - z_j) \in V_+ ,\;1\leq j\leq n-1 \} .
\label{tubular}
\end{equation}
In the case of the de Sitter space $X_d$ (embedded in ${{\Bbb R}}^{d+1}$), a
natural substitute for this property is to assume that $\WW_n$ is the boundary
value in the sense of distributions of a function holomorphic in 
\begin{equation}
\TT_n = X^{(c)n}_d \cap {\rm T}_n.
\label{tuboidal}
\end{equation}
It will be shown below that $\TT_n$ is a domain and a tuboid in the sense of 
\cite{[BM]}, namely a domain which is bordered by the reals
in such a way that the notion of ``distribution boundary value of a
holomorphic function from this domain'' remains meaningful.  
It is thus possible to impose:
\begin{enumerate}
\setcounter{enumi}{3}
\item (Weak spectral condition).
For each $n > 1$, the distribution ${\cal W}_{n}$ 
is the boundary value of a function ${\rm W}_{n}$ holomorphic in the subdomain
$\TT_n$ of $X^{(c)n}_d$.
\end{enumerate}
It may seem unnatural, in the absence of translational invariance,
to postulate analyticity properties in terms of the difference 
variables $( z_j - z_k)$. Note however that a Lorentz
invariant holomorphic function on a subdomain of $\Xcdn$ depends
only on the invariants $z_j\cdot z_k$. Among these the $z_j\cdot z_j$ are
fixed and equal to $-R^2$. Such a function therefore depends only
on the $( z_j - z_k)^2$. In the same way as in the 
Minkowskian case, it may be useful to relax some of the hypotheses
1-3. One may also want to impose:
\begin{enumerate}
\setcounter{enumi}{4}
\item (Temperedness Condition).
For each $n > 1$, there are constants $M(n) \ge 0$ and $L(n) \ge 0$
such that the distribution ${\cal W}_{n}$ 
is the boundary value of a function ${\rm W}_{n} $ holomorphic in the subdomain
$\TT_n$ of $X^{(c)n}_d$ satisfying
\begin{equation}
|{\rm W}_n(x+iy)| \le M(n)(1 + \|x+iy\| + 
{\rm dist}(z,\partial {\rm T}_n)^{-1})^{L(n)}.
\label{tempcond}
\end{equation}
\end{enumerate}
This global bound (which includes the behaviour of ${\rm W}_{n} $ at infinity) 
will not be indispensable in this paper, but the local part of it
(indicating a power behaviour near each point $ x $ for $ y $ tending to
zero) is in fact equivalent to the distribution character of the 
boundary value of ${\rm W}_{n}$ postulated  in 4 (see our remark 1 below).

For completeness, we now recall the definition of tuboids on  manifolds
(given in \cite{[BM]}).
Let ${\cal M}$ be a real $n$-dimensional analytic manifold, 
$T {\cal M} = \bigcup_{x\in {\cal M}} (x, T_x{\cal M})$ 
the tangent bundle to ${\cal M}$ and ${\cal M}^{(c)}$ a complexification of 
${\cal M}$. If $x_0$ is any point in ${\cal M}$, ${\cal U}_{x_0}$ and 
 ${\cal U}^{(c)}_{x_0}$ will denote open neighborhoods of $x_0$, respectively in ${\cal M}$ and ${\cal M}^{(c)}$ such that  
${\cal U}_{x_0}$ = ${\cal U}^{(c)}_{x_0}\cap {\cal M}$; a corresponding neighborhood of $(x_0,0)$ with basis  ${\cal U}_{x_0}$ in $T {\cal M}$
will be denoted $T_{{\rm loc}}{\cal U}_{x_0}$. 
\begin{definition}
\label{ald}
We call admissible local diffeomorphism at a point $x_0$ any diffeomorphism $\delta$ which maps some neighborhood 
$T_{{\rm loc}}{\cal U}_{x_0}$ of $(x_0,0)$ in  ${ T}{\cal M}$
onto a corresponding neighborhood ${\cal U}^{(c)}_{x_0}$ of $x_0$ in 
${\cal M}^{(c)}$ (considered as a $2n$-dimensional ${\cal C}^\infty$ manifold) in such a way that the following properties hold:
\begin{description}
\item{a)} $\forall x \in {\cal U}_{x_0}$, $\delta[(x,0)] =x$;
\item{b)} $\forall (x,y) \in {T}_{{\rm loc}}{\cal U}_{x_0}$, with  $y\not= 0,\, (y\in { T}_{x}{\cal M})$, the differentiable function
$t \to z(t) = \delta[(x,ty)]$
 is such that 
\begin{equation}
\frac{1}{i}\frac{dz}{dt}(t)|_{t=0} = \alpha y, \;\;\makebox{with} \;\;\alpha > 0.
\end{equation}
\end{description}
\label{a1}
\end{definition}

A tuboid can now be described as a domain in ${\cal M}^{(c)}$ 
which is bordered 
by the real manifold ${\cal M}$ and whose ``shape'' near each point 
of ${\cal M}$ is (in the space of $\Im z$ and for $\Im z \to 0$) 
very close to a given cone $\Lambda_x$ of the tangent space $T_x{\cal M}$ 
to  ${\cal M}$ at the point $x$. 
The following more precise definitions are needed.

\begin{definition} We call ``profile'' above ${\cal M}$ any 
open subset $\Lambda$ of ${ T}{\cal M}$ which is of the form 
$\Lambda= \bigcup_{x\in {\cal M}}(x,\Lambda_{x})$,
where each fiber 
$\Lambda_{x}$ is a non-empty cone with apex at the origin  
in ${   T}_{x}{\cal M}$ ($\Lambda_{x}$ can be the full tangent space  
${   T}_{x}{\cal M}$).\label{a2}
\end{definition}

It is convenient to introduce the ``projective representation''  
${ \dot{T}}{\cal M}$
of ${ T}{\cal M}$, namely  
${ \dot{T}}{\cal M} = \bigcup_{x\in {\cal M}}(x,{ \dot{T}}_{x}{\cal M})$, 
with ${ \dot{T}}_{x}{\cal M}$ = ${ {T}}_{x}{\cal M}\setminus \{0\}/
{{\Bbb R}}^{+}$. The image of  each point $y \in  { {T}}_{x}{\cal M}$ in 
${ \dot{T}}_{x}{\cal M}$ is  $\dot{y}=\{\lambda y;\,  \lambda >0\}$.
Each profile $\Lambda$ can then be represented by an open subset 
$\dot{\Lambda}= \bigcup_{x\in {\cal M}}(x,\dot{\Lambda}_{x})$ of  
${ \dot{T}}_{x}{\cal M}$ (each fiber $\dot{\Lambda}_{x} = {\Lambda}_{x}/
{{\Bbb R}}^{+}$ being now a relatively compact set).
We also introduce the complement of the closure of $\dot{\Lambda}$ in 
${ \dot{T}}{\cal M}$,
namely the open set 
$\dot{\Lambda}'$ =
${ \dot{T}}{\cal M}\setminus \overline{\dot{\Lambda}}$=
$\bigcup_{x\in {\cal M}}(x,\dot{\Lambda}'_{x})$ (note that 
$\dot{\Lambda}'_{x}\subset{ \dot{T}}_{x}{\cal M}
\setminus \overline{\dot{\Lambda}}_{x}$).
\begin{definition}
\label{ttuboids}
A domain $\Theta$ of ${\cal M}^{c}$ is called a tuboid with profile $\Lambda$ 
above ${\cal M}$ if it satisfies the following property.
For every point $x_0$ in $\cal M$, there exists an admissible local diffeomorphism $\delta$ at $x_0$ such that:
\begin{description}
\item{a)} every point $(x_0,\dot{y}_0)$ in $\dot{\Lambda}$ admits a compact neighborhood $K(x_0,\dot{y}_0)$ in $\dot{\Lambda}$ such that \\
$\delta \left[ \{(x,y);\, (x,\dot{y})\in K(x_0,\dot{y}_0),\,(x,y) \in 
{   T}_{{\rm loc}}{\cal U}_{x_0}\}\right]\subset \Theta$.
\item{b)} every point $(x_0,\dot{y}_0')$ in  $\dot{\Lambda}'$ admits a compact neighborhood $ K'(x_0,\dot{y}_0')$ in $\dot{\Lambda}'$ such that \\
$\delta\left[\{(x,y);\, (x,\dot{y})\in K'(x_0,\dot{y}_0),\,(x,y) \in 
{   T}_{{\rm loc}}{\cal U}'_{x_0}\}\right]\cap \Theta=\emptyset$
\end{description}   
In a) and b), $ T_{{\rm loc}}{\cal U}_{x_0} $ and
$ T_{{\rm loc}}{\cal U}'_{x_0} $ denote sufficiently small
neighbourhoods of $ (x_0, 0) $ in ${  T}{\cal M}$ which may depend
respectively on $ y_0 $ and $ y'_0 $, but always satisfy the conditions
of Definition \ref{ald} with respect to $\delta$. 

Each fiber ${\Lambda}_{x}$ of ${\Lambda}$ will also be called 
the profile at $x$ of the tuboid $\Theta$.
\end{definition} 

Using these notions and the results in appendix A of \cite{[BM]},
we will show the following

\begin{prop}
{i)} The set 
\begin{equation}
\TT_n=\{{ z} = ({ z}_{1},\ldots,{ z}_{n});\;{ z}_{k}=
{ x}_{k} + i{ y}_{k}\in X^{(c)}_{d},\;1\leq k\leq n; 
\;{ y}_{j+1}-
{ y}_{j}\in V^{+},\;1\leq j \leq n-1\}
\label{tubular1}
\end{equation}
is a domain of $\Xcdn$

{ii)} $\TT_n$ is a tuboid above  $\Xdn$, with profile
\begin{equation}
\Lambda^{n} = \bigcup_{\x\in \Xdn} (\x,\Lambda^{n}_{\x}) ,
\label{tubular3}
\end{equation}
where, for each $\x=(\x_{1},\ldots,\x_{n})\in \Xdn$, $\Lambda^{n}_{\x}$ 
is a non-empty open convex cone with apex at the origin in 
${ T}_{\x}\Xdn$ defined as follows:
\begin{equation}
{\Lambda}^{n}_{\x}=\{{ \y} = ({ \y}_{1},\ldots,{ \y}_{n});\;{ \y}_{k}\in 
{ T}_{\x_{k}}X_d,\;1\leq k \leq n; \;{ \y}_{j+1}-
{ \y}_{j}\in V^{+},\;1\leq j \leq n-1\}.
\label{tubular2}
\end{equation}
\label{pr1}
\end{prop}
\vskip5pt
{\bf Proof}
 \begin{description}
\item{a)} Let ${\rm C}_n$ be the open convex cone in $ {\Bbb R}^{nd} $
defined by 
\begin{equation}
{\rm C}_n =\{{ \y} = ({ \y}_{1},\ldots,{ \y}_{n});
\;{ \y}_{k}\in {\Bbb R}^{d},\;
1\leq k \leq n; \;{ \y}_{j+1}-
{ \y}_{j}\in V^{+},\;1\leq j \leq n-1\}
\end{equation}
The set ${\Lambda}^n $ defined in Eqs. (\ref{tubular3}) and (\ref{tubular2}) 
can then be seen as the restriction
of the open subset $\Xdn\times{\rm C}_n $ of $ \Xdn\times {\Bbb R}^{nd}$
to the algebraic set with equations $\x_j.\y_j = 0,\;1\leq j \leq n$,   
which represents $ T\Xdn $ as a submanifold of $\Xdn\times{\Bbb R}^{nd} $;
${\Lambda}^n$ is therefore an open subset of $ T\Xdn $. 
Moreover, for every point $\x=(\x_{1},\ldots,\x_{n})\in \Xdn$,
the set $\Lambda^{n}_{\x}$ defined in Eq. (\ref{tubular2}) is an open convex
cone in ${ T}_{\x}\Xdn$,as being the intersection of the latter with
$ {\rm C}_n $. For every $\x$, this cone is non-empty 
since one can determine at least one
vector $\y=(\y_{1},\ldots,\y_{n})\in\Lambda^{n}_{\x}$ as follows:
$\y_{1}$ being chosen arbitrarily in 
${T}_{\x_{1}}X_d$, we can always find 
$\y_{2}\in 
\left\{{T}_{\x_{2}}X_d\right\}\cap\left\{\y_{1}+V^{+}\right\}$,
 and then by recursion   
$\y_{j+1}\in 
\left\{{T}_{\x_{j}}X_d\right\}\cap\left\{\y_{j}+V^{+}\right\}$, 
for $j\leq n-1$, because
for every point $\x_{j}\in X_d$, 
$\left\{{T}_{\x_{j}}X_d\right\}\cap V^{+}$ is a non-empty convex
cone.
\item{b)} Let 
\begin{equation}
\Lambda^{n}_{R}=\bigcup_{\x\in\Xdn}(\x,\Lambda^{n}_{\x,R}) ,\;\;
\Lambda^{n}_{\x,R} = 
\{{ \y} = ({ \y}_{1},\ldots,{ \y}_{n})\in \Lambda^{n}_{\x} ,\;
{ \y}^{2}_{j}< R^{2},\;1\leq j \leq n\}.
\end{equation}
$\Lambda^{n}_{R}$ is (like $\Lambda^{n}$) an open subset of 
${T}\Xdn$;
each fiber  
$\Lambda^{n}_{\x,R}$ is a non-empty domain in  
${ T}_{\x}\Xdn$. This results from the property of
$\Lambda^{n}_{\x}$ proved in a), since the existence of a point in
$\Lambda^{n}_{\x,R}$ or of an arc connecting two arbitrary points inside
$\Lambda^{n}_{\x,R}$, follows from the corresponding property of
$\Lambda^{n}_{\x}$ by using the dilatation invariance of the latter.
It follows that $\Lambda^{n}_{R}$ is (like $\Lambda^{n}$) a connected
set and therefore a domain in 
${ T}\Xdn$.
\item{c)}
We now show that there exists a continuous mapping $\mu$  which is one-to-one from    
$ \Lambda^{n}_{R}$  onto the set $\TT_n\setminus Y^n_{R}$, where
$Y^n_{R}$ denotes the following subset of codimension $(d-1)$ of 
$\Xcdn$:    
\begin{equation}
Y^n_{R}=\{z=(z_{1},\ldots,z_{n};\;
z_{j}=x_{j}+iy_{j} \in X^{(c)}_d,\;1\leq j\leq n  
;\; \exists \;\makebox{at least one}\;
j_{0}: x_{j_{0}}=0 \; \}.
\end{equation}
 Let us consider the following mapping $\mu$:
\begin{equation}
\mu(\x,\y)=z=(z_{1},\ldots,z_{n}),\;\;z_{j}=x_{j}+iy_{j}=
\frac{\sqrt{R^{2}-\y^{2}_{j}}}{R}\x_{j}+i\y_{j},\;1\leq j\leq n.
\label{mu}
\end{equation}
$\mu$ is defined on the subset $ \left\{ T\Xdn  
\right\}_ R $ of all the elements $ ( \underline{x} , \underline{y}) $ of
$ T\Xdn $ such 
that $ \underline{y}^2_j < R^2 $ for $ 1 \leq  j \leq  n$; 
Eq. (\ref{mu}) 
implies that (for all $j$) $ z^2_j = -R^2 $ and therefore 
that $ \mu $ is a global 
diffeomorphism from $ \left\{ T\Xdn \right\}_ R
$ onto the subset $ Z^n_R= \Xcdn \setminus Y^n_R $ of $
\Xcdn; $ 
clearly, this diffeomorphism maps $ \Lambda^ n_R $ onto $ {\cal
T}_n\setminus Y^n_R, $ and therefore (in view of 
b)), $ \TT_n\setminus Y^n_R $ is a domain of $ \Xcdn$.
Since all points of $ \TT_n $ are either 
interior points or boundary points of $ \TT_n\setminus Y^n_R, $ and since $
\TT_n= \Xcdn\cap{\rm T}_n  $ 
is an open set, it is a domain of $ \Xcdn$.

\item{d)} In order to show that $ \TT_n $ is a tuboid 
with profile $ \Lambda^ n $ 
above $ \Xdn $,  
one just notices that the global diffeomorphism $ \mu $ provides 
admissible local
diffeomorphisms (by local restrictions) 
at all points $ \underline{x} $ in $ \Xdn $.  
Properties a) and b) of  Definition \ref{ttuboids}   are then satisfied by 
$\TT_n $ (with respect to all these local diffeomorphisms) 
as an obvious  
by-product of Eq. (\ref{mu}). 
\end{description}
 
\begin{remark}{\em By an application of theorem A.2. of \cite{[BM]}, the
weak spectral condition 
implies that for every $\x$ there is some local tube 
$\Omega_{\x}+i\Gamma_{\x}$ around $\x$ in any chosen system of local complex  coordinates on $\Xcdn$ whose image in  $\Xcdn$ 
is contained in $\TT_n$ has a profile very close to the profile of $\TT_n$ (restricted to a neighborhood of $\x$), from which the boundary value equation ${\cal W}_{n}=b.v.{\rm W}_{n}$ can be understood in the usual sense.
It implies equivalently that, in a complex neighborhood of each point $x=(x_{1},\ldots,x_{n})\in \Xdn$, the analytic function 
${\rm W}_{n}({z_{1}},\ldots,z_{n})$ is of moderate growth (i.e. bounded by a power of $\|y\|^{-1}$, where $\|y\|$ denotes any local norm of $ y=\Im z = (y_{1},\ldots, y_{n})$) when the point  $  z = (z_{1},\ldots, z_{n})$ tends to the reals inside $\TT_n$.}
\end{remark}

\begin{remark}
{\em An important difference with respect to the Minkowski case is that
the reals (i.e. $\Xdn$) are not a distinguished boundary for the tuboid $\TT_n$.}
\end{remark} 
\section{Consequences of locality, weak spectral condition and 
de Sitter covariance.}
Most of the well-known properties of the Wightman distributions 
in the Minkowskian case (\cite{[SW],[J]}) hold without
change in the de Sitterian case under our assumptions,
and their proofs mostly carry over literally. A few points,
however require some attention.
For each permutation $\pi$ of $(1,\ldots,n)$, the permuted Wightman 
distribution
\begin{equation}
{\cal W}^{(\pi)}_{n}(x_{1},\ldots,x_{n})=
{\cal W}_{n}(x_{\pi(1)},\ldots,x_{\pi(n)})
\end{equation}
is the boundary value of a function 
${\rm W}^{(\pi)}_{n}(z_{1},\ldots,z_{n})$  
holomorphic in the "permuted tuboid"
\begin{equation}
{\cal T}^{\pi}_n=\{{ z} = ({ z}_{1},\ldots,{ z}_{n});\;{ z}_{k}=
{ x}_{k} + i{ y}_{k}\in X^{(c)}_d,\;1\leq k \leq n;   
\;{ y}_{\pi(j+1)}-
{ y}_{\pi(j)}\in V^{+},\;1\leq j \leq n-1\}
\label{tubular11}
\end{equation}
If two permutations $\pi$ and $\sigma$ differ only by the exchange
of the indices $j$ and $k$, then $\WW_\pi$ and $\WW_\sigma$ 
coincide in
\begin{equation}
\RR_{j k} = X_d^n \cap {\rm R}_{jk},\ \ \ 
{\rm R}_{j k} = \{x \in \bR^{n(d+1)}\ :\ 
(x_j-x_k)^2 < 0\}.
\end{equation}
Let $\RR$ be a non-empty region which is the intersection of a subset of
$\{\RR_{j k}\ :\ j\not=k\}$.
By the edge-of-the-wedge theorem (in its version for tuboids, see
theorem A3 of \cite{[BM]}),
any maximal set of permuted Wightman distributions
which coincide on this region are the boundary value, in $\RR$,
of a common function holomorphic
in a tuboid above $\RR$ whose profile is 
obtained by taking {\em at each point} $x \in \RR$ 
the convex hull of the  
profiles at $x$ of the corresponding permuted tuboids.
In particular all the permuted Wightman distributions coincide in
the intersection $\Omega_n$ of all the $\RR_{j k}$, and it follows
that they all are boundary values
of a common function $\makebox{\Eul W}_{n}(z_{1},\ldots z_{n})$, 
holomorphic in a {\it primitive analyticity domain} ${\cal D}_{n}$.
$\makebox{\Eul W}_{n}$ is the common analytic continuation of all
the holomorphic functions ${\rm W}^{(\pi)}_{n}$ 
and the domain ${\cal D}_{n}$ is the union of all the 
permuted tuboids ${\cal T}^{\pi}_{n}$ and of the above mentioned local
tuboids associated (by the edge-of-the-wedge theorem) 
with finite intersections
of the $\RR_{j k}$. In particular ${\cal D}_{n}$ contains a complex
neighborhood of $\Omega_n$ since the tuboids 
$\TT^{\pi}_n$ and ${\TT^{\pi_{inv}}}_n$ are opposite (where
$\pi_{inv} = (\pi(n),\ldots,\ \pi(1))$). For each permutation
$\pi$ we denote $\TT^{\pi\ \rm ext}_n$ the extended permuted tuboid
\begin{equation}
\TT^{\pi \rm ext}_n = \bigcup_{\Lambda_c \in L_+(\bC)} \Lambda_c \TT^{\pi}_n =
\bigcup_{\Lambda_c \in L_+(\bC)} \Lambda_c({\rm T}^{\pi}_n\cap \Xcdn) =
\Xcdn\cap \bigcup_{\Lambda_c \in L_+(\bC)} \Lambda_c {\rm T}^{\pi}_n =
\Xcdn\cap {\rm T}^{\pi \rm ext}_n.
\end{equation}

\subsection
{The Jost points and the Glaser-Streater theorem}
The set of real points of ${\rm T}^{\rm ext}_n = {\rm T}^{1\, \rm ext}_n$
(Jost points in the ambient space) is denoted ${\rm J}_n$. Its intersection 
$\JJ_n$ with
$X_d^n$ will be called 
{\em the set of Jost points associated with the tuboid $\TT_n$}.
The set $\JJ_n $ is generated (like ${\rm J}_n$)  by the action 
of the connected group $\Gd$ on a special subset  
of Jost points 
associated with a given maximal space-like cone
such as the ``right-wedge'' $ W_{(r)} $ of the ambient space: 
\begin{equation}
W_{(r)} = - W_{(l)} = \{x \in \bR^{d+1}\ :\ u(x) >0,\ \ v(x)<0\},
\end{equation} 
the notations $u$,$v$ being those of Eq. (\ref{udef}).
The corresponding {\em special Jost subset} ${\JJ_n}^{(r)}$ 
is defined by 
\begin{equation}
{\JJ_n}^{(r)} = {\rm J}_n^{(r)} \cap X_d^n, 
\end{equation}
with
\begin{equation}
{\rm J}_n^{(r)} = 
\{(x_1,\ \ldots ,x_n) \in \bR^{n(d+1)}\ :\ 
x_1 \in W_{(r)},\ (x_2 - x_1) \in W_{(r)},\ \ldots ,\ (x_n-x_{n-1}) \in W_{(r)}\}. 
\end{equation} 
The fact that
$\JJ_n$ 
is a non-empty and, if $d >2$, connected set 
is then a consequence of the connectedness of 
${\JJ}_n^{(r)} $.
The latter property can be checked as follows. 
The projection
${[{\JJ}_n^{(r)}]}_{u,v} $ of 
${\JJ}_n^{(r)} $
onto the space ${\Bbb R}^{2n}$ of the $(u,v)$-coordinates
is the intersection of the convex cone $ (u_1 >0,\ v_1 <0,\ 
u_{j+1} - u_{j} >0,\ v_{j+1} - v_{j} <0,\ 1\leq j \leq n-1) $ 
(here we have put $u_j = u(x_j), v_j = v(x_j) $) 
with the set $ (u_1v_1 > - R^2, \ldots, u_nv_n > - R^2)$ 
which is preserved by the contractions; therefore, any couple of points in
${[{\JJ}_n^{(r)}]}_{u,v} $ can be connected by a broken line
contained in this set. Considering now 
${\JJ}_n^{(r)} $ as a fiberspace over its projection 
${[{\JJ}_n^{(r)}]}_{u,v} $, we see that it is locally trivialized
with a toroidal fiber of the form $ {\underline{\vec{x}}}_j^2 = constant,
\ 1\leq j \leq n $ which is connected provided $d$ is larger than $2$;
the connectedness of 
${\JJ}_n^{(r)} $ follows correspondingly. 
\vskip 1cm

As in the Minkowskian case, one can then state a de Sitterian version
of the Glaser-Streater property, according to which
any function holomorphic in $\TT_n\cup -\TT_n\cup{\JJ}_n$ 
has a single-valued 
analytic continuation in $\TT_n^{\rm ext} = \TT_n^{1\ \rm ext}.$
(see e.g. \cite{[BEG],[J],[St]}).
Hence every permuted Wightman distribution is the boundary value of
a function holomorphic in the corresponding extended permuted tuboid
$\TT^{\pi\ \rm ext}_n$;   
this function is in fact an analytic continuation of ${\rm W}^{(\pi)}_{n}$
and thereby of the common holomorphic $n$-point function
$\makebox{\Eul W}_{n} $.   
\vskip 1cm

\noindent {\bf Remark.} 
The proof of the Glaser-Streater property is based on a lemma of
analytic completion in the orbits of the complex Lorentz group
and this is why it holds for the complexified 
de Sitter space (since $\Xcdn $ is a union 
of such orbits), the connectedness of the set of orbits 
generated by the Jost points being of course crucial. 
To be complete, one must also point out that it
requires the following strong form of the Bargmann-Hall-Wightman lemma,
(\cite{[HW]}, pp. 95-97, \cite{[SW]} pp. 67-70) proved for $d+1 \le 4$ in
these references, and extended to all dimensions in \cite{[J1]}.
\ifundefined{BHWextApp}{}\else
An alternative proof of the latter  
is given in Appendix \ref{BHW}.\fi

\begin{lemma}
(Bargmann-Hall-Wightman-Jost)
Let $M\in L_+(\bC)$ be such that 
${\rm T_+}\cap M^{-1}{\rm T_+} \not= \emptyset$. There exists a 
continuous map $t \mapsto M(t)$ of $[0,\ 1]$ into $L_+(\bC)$
such that $M(0) =1$, $M(1) = M$, and that, for every 
$z \in {\rm T_+}\cap M^{-1}{\rm T_+}$ and $t\in[0,\ 1]$,
$M(t)z \in {\rm T_+}$.
\end{lemma}

\subsection{The PCT-property}
The standard proof of the PCT theorem (see \cite{[J],[SW]} and
references therein) extends in a straightforward 
way to the de Sitterian case
under the assumptions of covariance, weak spectral condition, and
locality. The latter can be relaxed to the condition of weak locality
\cite{[Dy],[J],[SW]}, namely:

\noindent{\bf Weak locality condition:} {\it For every Jost point
$(r_1,\ldots,\ r_n) \in \JJ_n$,}
\begin{equation}
\WW_n(r_1,\ldots,\ r_n) = \WW_n(r_n,\ldots,\ r_1)
\end{equation}
which obviously follows from locality.

\begin{prop}(PCT invariance)
{}From the weak spectral condition, the covariance condition, and
the weak locality condition, it follows that
\begin{equation}
{\cal W}_{n}(x_{1},\ldots,x_{n})=
{\cal W}_{n}({\rm  I}_0 x_{n},\ldots, {\rm  I}_0x_{1})
\label{pct}
\end{equation}
holds at every real $x\in\Xdn$ (in the sense of distributions),
where ${\rm  I}_0 = -1$ if $d$ is odd, and, if $d$ is even, for
every $z\in\bC^{d+1}$, 
\begin{equation}
({\rm  I}_0\,z)^{(\mu)} = - z^{(\mu)}\ \ {\rm for\ }
0\le\mu <d,\ \ \ ({\rm  I}_0\,z)^{(d)} = z^{(d)}.
\end{equation}
If moreover the positivity condition holds, there exists an 
antiunitary operator $\Theta :\ \HH \rightarrow \HH$ such that
\begin{equation}
\Theta \Omega = \Omega, \ \ \ 
\Theta \langle\Phi_n^{(b)},\ f \rangle =
\langle\Phi_n^{(b)},\ f^\star_{{\rm  I}_0} \rangle ,
\end{equation}
where 
$f^\star_{{\rm  I}_0}(x_1,\ldots,\ x_n) =
\bar f({\rm  I}_0 x_{n},\ldots, {\rm  I}_0x_{1})$.
\end{prop}

One notices that, in this statement, it is 
the symmetry ${\rm I}_0 $ (which depends on the parity of the dimension) 
which has to be used (as it is also 
the case for $ d+1 $-dimensional Minkowskian theories). This is
due to the fact that ${\rm I}_0$ always belongs to the corresponding
complex connected group $ L_+(\Bbb C) $ under which the
functions $ \makebox{\Eul W}_n $ are invariant. Since the mapping
$ (z_1,\ldots,z_n) \to ({\rm I}_0 z_n, \ldots,{\rm I}_0 z_1) $ 
is always (for every $n$) an
automorphism of the tuboid $ {\cal T}_n $, the standard analytic
continuation argument \cite{[J],[SW]} applies to the proof of 
Eq. (\ref{pct}). Now, it is interesting to note that for $d$ even
(in particular in the ``physical case'' $ d=4$)
${\rm I}_0 $ does have the interpretation of a {\em space-time inversion} 
in a local region of the de Sitter universe
around the {\em base point} $ x_0 $ with coordinates
$ (0,\ldots,0,R)$, considered as playing the role of 
the origin in Minkowski space. In fact,   
the stabilizer of $x_0$ 
(inside the de Sitter group) is 
the analogue of the Lorentz group 
(inside the Poincar\'e group) and indeed    
it acts as the latter in the (Minkowskian) tangent
space to $ X_d $ at $x_0$; 
${\rm I}_0$ then appears as the corresponding
space-time inversion   
(contained in the complexified stabilizer of $x_0$ ).
This means that (for $d$ even) the previous proposition can be seen as 
introducing a {\em PCT-symmetry relative to the point $x_0$};
analogous symmetry operators could be associated with
all points of the de Sitter manifold.

\subsection {Euclidean points}
In the ambient complex $n$-point Minkowskian space-time $\bC^{n(d+1)}$,
the union of the permuted extended tubes 
$\bigcup_\pi  {\rm T}^{\pi \rm ext}_n$ contains all non-coinciding
Euclidean points. Since the intersection of this union with $\Xcdn$ is
the union of all permuted extended tuboids $\TT^{\pi \rm ext}_n$,
it follows that the domain of analyticity of $\makebox{\Eul W}_{n}$ 
contains the set of all non-coinciding points of the product of
$n$ Euclidean spheres.

\subsection{The case $n = 2$. Generalized free fields
and their Wick powers}
The extended tube ${\rm T}^{\rm ext}_2$ is equal to
$\{(z_1,\ z_2) \in \bC^{2(d+1)}\ :\ (z_1 -z_2)^2 \notin \Rpc\}$.
Hence
\begin{equation}
\TT^{\rm ext}_2 = 
\{(z_1,\ z_2) \in X_d^{(c)2}\ :\ (z_1 -z_2)^2 \notin \Rpc\}.
\end{equation}
In particular 
$\makebox{\Eul W}_{2}(z_1,z_2)-\makebox{\Eul W}_{2}(z_2,z_1)$ 
is analytic, odd, and
Lorentz invariant at real space-like separations, hence vanishes there
even without the locality assumption.
Thus under the assumptions of weak spectral condition and covariance,
${\rm W}_2(z_1,\ z_2)$ defines an ``invariant perikernel'' in the sense
of \cite{[BV-2]} which can be represented by a function 
$w({\zeta}) $ 
of the single complex variable
${\zeta} = 1+(z_1 -z_2)^2/2R^2 = -z_1\cdot z_2/R^2$, 
holomorphic in the cut-plane $\bC \setminus [1,\ \infty)$.  
Any such two-point function completely
determines a generalized free field $A$ whose Wightman functions
are obtained by the same formulae as in the Minkowskian case. 
(see \cite{[BM]} for a detailed study of all that). 
$A$ can also be seen as the restriction of
a generalized free field on the ambient Minkowski space, in general
with an indefinite metric (see also in this connection subsection 5.4 of
\cite{[BM]}).  
Wick monomials in $A$ have well-defined Wightman functions, again given
by the same formulae as in the Minkowskian case, i.e. as sums of products
of two-point functions. Since these Wightman functions can be obtained
as limits of Wightman functions of Wick monomials of group-regularizations
of $A$, they satisfy all the conditions 1-5 (in particular positivity)
provided $A$ does. In
particular the Wick monomials in $A$ are unbounded distribution valued
operators in the Fock space of $A$, and provide examples of theories 
satisfying all the axioms.

\section{Physical interpretation of the weak spectral condition} 
 
In this section,
we are still in the Lorentz coordinate frame $\{e_0,\ldots,\ e_d\}$ in
the ambient real Minkowski space, the notations $u$, $v$, $[\lambda]$ are 
as in Eqs. (\ref{udef}) and (\ref{speclor}).

Let us now discuss the physical interpretation of the spectral condition 
we have introduced.  Following the pioneering approach of Unruh \cite{[U]}, 
Gibbons and Hawking \cite{[GH]} we  adopt the viewpoint of a geodesical 
observer and namely the one   moving on the geodesic $h(x_0)$
of the base point $x_0$ contained in the $(x^{(0)},x^{(d)})$-plane, 
which we parametrize as follows: 
\begin{equation}
h(x_0)= \{x=x(\tau);\;\;\;x^{(0)}=R\sinh \frac{\tau}{R},\;
x^{(1)}=\cdots= x^{(d-1)}=0,\;x^{(d)}=R\cosh\frac{\tau}{R}\}
\label{geod}\end{equation}
The parameter $\tau$ of the representation (\ref{geod}) 
is   the proper
time of the observer and the base point $x_0$ is the event for 
which ${\tau} = 0$ .

The set of all events of  $X_d$ which
can be connected  with the observer by the reception and the emission of   
light-signals is the region: 
\begin{equation}
{\cal U}_{h(x_0)} = \{x \in X_d: \;\;\;x^{(d)}>|x^{(0)}|\} = 
W_{(r)}\cap X_d.
\end{equation}
Points in   ${\cal U}_{h(x_0)}$ can be parametrized by 
$(\tau, \vec{\x})$ as follows:
\begin{equation}
x(\tau,\vec{\underline{x}})  
=
\left[
\begin{array}l
x^{(0)} =  \sqrt{R^2 - \vec{\underline{x}}^2 }\sinh  \frac{\tau}{R} \\
( x^{(1)},\ldots,x^{(d-1)}) = \vec{\underline{x}}\\
x^{(d)}=\sqrt{R^2- \vec{\underline{x}}^2 } \cosh  \frac{\tau}{R}  
\end{array}
\right. , \;\;\;\;\;\tau\in {\Bbb R},\;\;{\vec{ \x}}^2<R^2
\label{parahor}
\end{equation}
${\cal U}_{h(x_0)}$ 
is the intersection of the hyperboloid with the wedge $W_{(r)}$ 
of the ambient space
and  admits two boundary parts $H^+_{h(x_0)}$ and $H^-_{h(x_0)}$,
respectively called the ``future'' and ``past horizons'' of the geodesical 
observer:
\begin{equation}
H^\pm_{h(x_0)} = \{x \in X_d: \;\;\;x^{(0)}=\pm x^{(d)},\;\;x^{(d)} \ge 0\}.
\end{equation}
${\cal U}_{h(x_0)}$ is stable under  the transformation (\ref{speclor}), 
for $\lambda = e^{\frac{t}{R}} > 0$.  These transformations constitute a 
subgroup $T_{h(x_0)} $   of $\Gd$. The  action of  $ T_{h(x_0)}(t)$ on 
${\cal U}_{h(x_0)}$ written in terms of the parameters $t$ and $\tau$    
can be interpreted as a ``time-translation'':
\begin{equation}
T_{h(x_0)}(t) [x(\tau,\vec{\underline{x}})]  
=x(t+\tau,\vec{\underline{x}})\equiv x^ {t} .
\label{old}
\end{equation}
  $T_{h(x_0)}$ thus defines a group of 
isometric automorphisms of 
${\cal U}_{h(x_0)}$ whose orbits are all branches of hyperbolae of 
${\cal U}_{h(x_0)}$ in two-dimensional plane sections parallel to the 
$(x^{(0)},x^{(d)})$-plane (see \cite{[KW]} for a general 
discussion of this kind of structure). 
 
Before discussing the physical interpretation of the spectral condition, 
we need to extend to the de Sitter case 
one aspect of a well-known result of Bisognano and Wichmann [BW] 
which concerns analyticity properties in orbits of the complexified group 
$ T_{h(x_0)}^{(c)}$ of $T_{h(x_0)}$.

\subsection{ Bisognano-Wichmann analyticity.}

For every function $g_n$ in $\DD(\Xdn)$ or $\SS(\Xdn)$ and every
$\lambda \in \bR\setminus \{0\}$, $[\lambda]$  as in  Eq. (\ref{speclor}), we denote
(with a simplified form of (\ref{Lambdaact}))
\begin{equation}
{g_n}_\lambda(x_1,\ldots,\ x_n) = 
g_n([\lambda^{-1}]x_1,\ldots,\ [\lambda^{-1}]x_n).
\end{equation}
and
\begin{equation}
g_n^{\leftarrow}(x_1,\ldots,\ x_n) = 
g_n(x_n,\ldots,\ x_1)   
\end{equation}

Then one has:

\begin{thm}
\label{bw}
If a set of Wightman distributions satisfies the locality and weak
spectral conditions, then for all $m,\ n\in \bN$, 
$f_m\in \DD(W_{(r)}^m\cap X_d^m)$ and $g_n\in \DD(W_{(r)}^n\cap X_d^n)$,
there is a function $G_{(f_m,g_n)}(\lambda)$ holomorphic on 
$\bC\setminus {\Rpc}$ 
with continuous boundary values $G_{(f_m,g_n)}^\pm$
on $\Rpo$ from the upper and lower half-planes such that: 

a) for all $\lambda \in \Rpo$,
\begin{equation}
G_{(f_m,g_n)}^+(\lambda) =
\langle\WW_{m+n},\ f_m\otimes {g_n}_\lambda\rangle,\ \ \ \ 
G_{(f_m,g_n)}^-(\lambda) = 
\langle\WW_{m+n},\ {g_n}_\lambda\otimes f_m\rangle.
\label{defG}
\end{equation}

b) for all $\lambda \in \Rmo$,
\begin{equation}
G_{(f_m,g_n)}(\lambda) =
\langle\WW_{m+n},\ f_m\otimes {g_n^{\leftarrow}}_\lambda\rangle =  
\langle\WW_{m+n},\ {g_n^{\leftarrow}}_\lambda\otimes f_m\rangle. 
\label{Gantipod}
\end{equation} 
\end{thm}

This theorem neither requires positivity nor Lorentz covariance.
It expresses a property of the domain of holomorphy of the
Wightman functions, and of the boundary values from this domain.
In fact, it states that appropriate boundary values of the $(m+n)-$point
holomorphic function ${\W}_{m+n}$, taken in the region where
all the variables $w_1,\ldots,\ w_m,\ z_1,\ldots,\ z_n$ belong to 
$W_{(r)}\cap X_d$, are holomorphic with respect to the group variable
$\lambda$ (for $\lambda \in {\bC}\setminus \Rpc$) in the 
orbits $ (w,x) \mapsto (w,[\lambda]x)$ of $T_{h(x_0)}^{(c)}$  
(with $w =(w_1,\ldots,\ w_m),\ \ x = (x_1,\ldots,\ x_n),
\lambda = e^{t\over R}$) and such that:
\vskip 12pt

for $\lambda > 0$,
\begin{equation}
{\W}_{m+n} (w,\  [\lambda + i 0] x) 
= {\rm W}_{m+n} (w,
\ [\lambda]x),\ \ \  
{\W}_{m+n} (w,\  [\lambda - i 0] x) 
= {\rm W}_{m+n} 
([\lambda] x,\ w)
\label{a)}
\end{equation}

and for $\lambda < 0$, putting $x_{\leftarrow} = (x_n,\ldots,\ x_1)$, 
\begin{equation}
 \W_{m+n} (w,\ [\lambda]x) =
{\rm W}_{m+n}(w,\ [\lambda]x_{\leftarrow}) =
{\rm W}_{m+n}([\lambda]x_{\leftarrow},\ w) 
\label{c)}
\end{equation}
the latter equality being a direct consequence of locality
(since $ x \in W_{(r)}^n $ and $\lambda < 0$ imply $[\lambda]x_{\leftarrow}
\in W_{(l)}^n$). 
 
The  theorem will be proved here under the simplifying assumption
that the temperedness condition (\ref{tempcond}) holds.
\vskip 12pt

\noindent{\bf Proof}\HB
Four permuted branches of the
function ${\W}_{m+n}$
are involved in the proof. The variables $w=(w_1,
\ldots,\  w_m)$ will always be kept real in  
$W_{(r)}^m\cap X_d^m$, while the variables $z = (z_1,\ldots,\ z_n)$ are 
complex (in $\Xcdn$)
and we denote $y={\rm Im} \, z$.
The corresponding analyticity domains in the variables $z$
(described below) are obtained in the boundaries 
(i.e. in the ``face'' $ \Im w = 0$) of four 
permuted tuboids ${\cal T}^{\pi}_{m+n}$ 
according to the prescription of our weak spectral condition.
In view of the distribution boundary value procedure,
restricted to the subset of variables $w$, these analyticity domains
are obtained whenever one smears out the permuted functions ${\rm W}_{m+n}^{\pi}$ 
under consideration with a fixed function $ f_m \in \DD(W_{(r)}^m\cap X_d^m)$. 
(this function being understood as \it the \rm function named $f_m$  
in the statement of the theorem). These four branches are:

\begin{enumerate}
\item[i)] ${\rm W}_{m+n} (w_1,\ldots,\ w_m ,\  z_1
,\ldots,\  z_n) = 
{\rm W}_{m+n} (w,\ z)$,    
holomorphic in the tuboid:\\
${\cal Z}_{n\, +} = \left\{ z\in {\Xcdn};\  
y_1 \in V_+ , y_j - y_{j-1} \in V_+ ,\,j = 2,\ldots,\  n  \right \}$;
\item[ii)] ${\rm W}_{m+n} (z_n,\ldots,\ z_1,\ w_1, 
\ldots,\  w_m) =  
{\rm W}_{m+n} (z_{\leftarrow},\ w)$,    
holomorphic in the opposite tuboid:\\
${\cal Z}_{n\, -} = \{z \in {\Xcdn};\  y_1 \in V_- , y_j - y_{j-1} \in V_-,\  
j=2,\ldots,\ n \}$; 
\item[iii)] ${\rm W}_{m+n} (z_1,\ldots,\ z_n,\ w_1
,\ldots,\  w_m) =   
{\rm W}_{m+n} (z,\ w)$,    
holomorphic in the tuboid:\\
 ${\cal Z}'_{n\, +} = \{z \in {\Xcdn};\   y_n
\in V_- , y_j - y_{j-1} \in V_+ , j=2, \ldots ,n \}$;
\item[iv)] ${\rm W}_{m+n} (w_1,\ldots,\ w_m,\ z_n
,\ldots,\ z_1) =   
{\rm W}_{m+n} (w,\ z_{\leftarrow})$,    
holomorphic in the opposite tuboid:\\
 ${\cal Z}'_{n\, -} = \{z \in {\Xcdn};\  
y_n \in V_+ , y_j - y_{j-1} \in V_- ,j=2\ldots,\  n
\}$.
\end{enumerate}

Correspondingly, 
with the fixed function $ f_m \in \DD(W_{(r)}^m\cap X_d^m)$
we associate the following four functions
$z \mapsto F_{\pm}({f_m};\ z)$ and $z \mapsto F'_{\pm}({f_m};\ z)$: 
\begin{equation}
F_+({f_m};\ z) = \int_{\Xdm} {\rm W}_{m+n}(w,\ z)\  
f_m(w)\ 
d^m \sigma(w),\ \ \ \ \  
F_-({f_m};\ z) = \int_{\Xdm}  
{\rm W}_{m+n} (z_{\leftarrow},\ w)\     
f_m(w)\ 
d^m \sigma(w)
\label{F}
\end{equation}
\begin{equation}
F'_+({f_m};\ z) = \int_{\Xdm}  
{\rm W}_{m+n} (z,\ w)\    
f_m(w)\ 
d^m \sigma(w),\ \ \ \ \     
F'_-({f_m};\ z) = \int_{\Xdm}   
{\rm W}_{m+n} (w,\ z_{\leftarrow})\     
f_m(w)\ 
d^m \sigma(w)
\label{F'}
\end{equation}
which are  respectively
holomorphic in ${\cal Z}_{n\, +}$, ${\cal Z}_{n\, -}$, ${{\cal Z}'}_{n\, +}$ and ${{\cal Z}'}_{n\, -}$.  
By letting the variables $z$ tend to the reals 
from the respective tuboids 
${\cal Z}_{n\, +}$, ${\cal Z}_{n\, -}$, ${{\cal Z}'}_{n\, +}$ and ${{\cal Z}'}_{n\, -}$, 
and taking the corresponding boundary values 
$F_{\pm}^{(b)}({f_m};\ x)$ 
and ${F'}_{\pm}^{(b)}({f_m};\ x)$ 
of $F_{\pm}$ and $F'_{\pm}$ 
on $\Xdn$ in the sense of
distributions, one then obtains 
for every $g_n\in\DD(\Xdn)$
the following relations which involve the $(m+n)-$point Wightman distributions
considered in the statement of the theorem:  
\begin{equation}
\int_{\Xdn} F_+^{(b)}({f_m};\ x)\,{g_n}(x)\,
d^n \sigma(x) \ =\ \  
\langle\WW_{m+n},\ f_m\otimes {g_n}\rangle,\ \ \ \ 
\label{bv1}
\end{equation}
\begin{equation}
\int_{\Xdn} F_-^{(b)}({f_m};\ x)\,{g_n}(x)\,
d^n \sigma(x) \ =\ \  
\langle\WW_{m+n},\ {g_n^{\leftarrow}}\otimes f_m\rangle. 
\label{bv2}
\end{equation}

\begin{equation}
\int_{\Xdn} {F'}_+^{(b)}({f_m};\ x)\,{g_n}(x)\,
d^n \sigma(x) \ =\ \  
\langle\WW_{m+n},\ {g_n}\otimes f_m\rangle.
\label{bv3}
\end{equation}
\begin{equation}
\int_{\Xdn} {F'}_-^{(b)}({f_m};\ x)\,{g_n}(x)\,
d^n \sigma(x) \ =\ \  
\langle\WW_{m+n},\ f_m\otimes {g_n^{\leftarrow}}\rangle .  
\label{bv4}
\end{equation}

We now notice that, in view of local commutativity,  
$F_+^{(b)}({f_m};\ x)$ and $F_-^{(b)}({f_m};\ x)$ coincide  
in the sense of distributions on the set of special Jost points 
$\JJ_n^{(l)} = -\JJ_n^{(r)} = \{(x_1,\ldots,\ x_n) \in \Xdn;\   
0 > u_1 >\ldots u_{n-1} > u_n,\ \ 0 < v_1 <\ldots v_{n-1} < v_n \}$; 
therefore, in view of 
the edge-of-the-wedge theorem, the functions  
$z\mapsto F_+({f_m};\ z)$ and $z\mapsto F_-({f_m};\ z)$ have a common holomorphic
extension, denoted $F({f_m};\ z)$,   
in $\Delta = {\cal Z}_{n\, +} \cup {\cal Z}_{n\, -} \cup \VV$, where 
$\VV$ is a complex neighborhood of $\JJ_n^{(l)}$, such that
$[\lambda]\VV = \VV$ for all $\lambda >0$ (in particular 
$F_+^{(b)}({f_m};\ x)$ and $F_-^{(b)}({f_m};\ x)$ are continuous on $\JJ_n^{(l)}$).
By a similar use of local commutativity for ${F'}_+^{(b)}$ and ${F'}_-^{(b)}$,   
which coincide on the set of special Jost points
${\JJ'}_n^{(l)} = 
\{(x_1,\ldots,\  
x_n) \in \Xdn;  0 > u_n > u_{n-1} \ldots
> u_1, \ \ \ 0 < v_n < v_{n-1}\ldots < v_1\}$,  
we also notice that the functions  
$z\mapsto {F'}_+({f_m};\ z)$ and $z\mapsto{F'}_-({f_m};\ z)$ have a common holomorphic
extension, denoted $ F'({f_m};\ z)$, 
in ${\Delta'} = {{\cal Z}'}_{n\, +} \cup {{\cal Z}'}_{n\, -} \cup {\VV'}$, where 
${\VV'}$ is a complex neighborhood of ${\JJ'}_n^{(l)}$, such that
$[\lambda]{\VV'} = {\VV'}$ for all $\lambda >0$.  
Moreover, if the temperedness condition (\ref{tempcond}) is satisfied by the function
${\rm W}_{m+n}$, it can be checked that  
similar inequalities are satisfied by the holomorphic   
functions $F(f_m;\ z)$ and $F'(f_m;\ z)$ with respect to the variables $z$
in their respective tuboids ${{\cal Z}}_{n\, \pm}$ and ${{\cal Z}'}_{n\, \pm}$.

At this point, we shall rely on the following basic lemma 
which provides analytic completion in the orbits 
of the group $\{z \mapsto [\lambda]z \}$ (for $\lambda \in \bC_{\pm}$) and  
whose proof is given below (after the end of our argument). 

\begin{lemma}
\label{analcomp} 
a)\ \  Given any function $H(z)$ holomorphic in $\Delta$, the function
$(z,\ \lambda) \mapsto H([\lambda]z)$ is holomorphic in 
${\cal Z}_{n\, +} \times \bC_+$. Moreover, if $H(x + iy)$ satisfies            
majorizations of the form (\ref{tempcond}) in the tuboids
${\cal Z}_{n\, +}$ and ${\cal Z}_{n\, -}$ allowing one to define   
the boundary values $ H_+^{(b)}$ and
$H_-^{(b)}$ of $H$ from ${\cal Z}_{n\, +}$ and ${\cal Z}_{n\, -}$
as tempered distributions,
then the function  
$(z,\ \lambda) \mapsto H([\lambda]z)$ admits a
distribution boundary value on 
$\Xdn \times \bC_+$           
(still denoted 
$H([\lambda]x)$); the latter is a distribution in $x$ with values in the  
functions of $\lambda$ which are holomorphic in 
$ \bC_+$ and continuous in
$ \overline {\bC_+} \setminus \{0\}$ and one has:  
\begin{equation}
H([\pm\lambda]x) = H_{\pm}^{(b)}([\pm\lambda]x) \ \ \ \
{\rm for}\ \lambda >0  
\label{BV1}
\end{equation}
(the latter being identities between distributions in $x$
with values in the continuous functions of $\lambda$).

b)\ \ Similarly, given any function $H'(z)$ holomorphic in ${\Delta}'$,   
the function $(z,\ \lambda) \mapsto H'([\lambda]z)$ is holomorphic
in $\ZZ'_+ \times \bC_-$. Moreover, if $H'(x + iy)$ satisfies            
majorizations of the form (\ref{tempcond}) in the tuboids
$\ZZ'_+$ and $\ZZ'_-$ allowing one to define   
the boundary values $ {H'}_+^{(b)}$ and
${H'}_-^{(b)}$ of $H'$ from $\ZZ'_+$ and $\ZZ'_-$
as tempered distributions,
then the function  
$(z,\ \lambda) \mapsto H'([\lambda]z)$ admits a
distribution boundary value on 
$\Xdn \times \bC_-$,           
holomorphic in 
$ \bC_-$ and continuous in
$ \overline {\bC_-} \setminus \{0\}$, and one has:  
\begin{equation}
H'([\pm\lambda]x) = {H'}_{\pm}^{(b)}([\pm\lambda]x)
\ \ \ \ {\rm for}\ \lambda >0.  
\label{BV2}
\end{equation}
\end{lemma}

Since the function $F({f_m};\ z)$ satisfies the analyticity and temperedness
properties of the function $H(z)$ of Lemma \ref{analcomp} a), it follows that
one can take the boundary value onto $\Xdn \times \bC_+$ from 
${\cal Z}_{n\, +} \times \bC_+$ of the holomorphic  
function $(z,\ \lambda) \mapsto  F({f_m};\ [\lambda]z)$ 
and obtain for every $g_n \in {\cal D}(\Xdn)$ the 
following relations (deduced from Eq. (\ref{BV1}) after taking into account
Eqs. (\ref{bv1}) and (\ref{bv2})): 
\begin{equation}
\int_{\Xdn}\ F({f_m};\ [\lambda]x) g_n(x) d^n\sigma(x) 
= \langle\WW_{m+n},\ f_m\otimes {g_n}_{\lambda} \rangle \ \ \ \ {\rm for}\ \lambda > 0, \label{bvg1}
\end{equation}
\begin{equation}
\int_{\Xdn}\ F({f_m};\ [\lambda]x) g_n(x) d^n\sigma(x) 
= \langle\WW_{m+n},\ {g_n^{\leftarrow}}_{\lambda}\otimes f_m\rangle\  
\ \ \ \ {\rm for}\ \lambda < 0. 
\label{bvg2}
\end{equation}

Similarly, one can apply the results of 
Lemma \ref{analcomp} b)
to the function $H'(z) = F'({f_m};\ z)$; 
one can thus take the boundary value onto $\Xdn \times \bC_-$ from 
${{\cal Z}'}_{n\, +} \times \bC_-$ of the holomorphic  
function $(z,\ \lambda) \mapsto  F'({f_m};\ [\lambda]z)$ 
and obtain for every $g_n \in {\cal D}(\Xdn)$ the 
following relations (deduced from Eq. (\ref{BV2}) after taking into account
Eqs. (\ref{bv3}) and (\ref{bv4})): 

\begin{equation}
\int_{\Xdn}\ F'({f_m};\ [\lambda]x) g_n(x) d^n\sigma(x) 
= \langle\WW_{m+n},\ {g_n}_{\lambda}\otimes f_m\rangle\ \ \ \ {\rm for}\ \lambda > 0, \label{bvg'1}
\end{equation}
\begin{equation}
\int_{\Xdn}\ F'({f_m};\ [\lambda]x) g_n(x) d^n\sigma(x) 
= \langle\WW_{m+n},\ f_m\otimes {g_n^{\leftarrow}}_{\lambda}\rangle \ \ \ \ 
{\rm for}\ \lambda < 0  . 
\label{bvg'2}
\end{equation}

The l.h.s. of Eqs. (\ref{bvg1}) (or   (\ref{bvg2})) 
and (\ref{bvg'1}) (or (\ref{bvg'2}))
are respectively the boundary values of the holomorphic functions    
\begin{equation}
G_{(f_m,g_n)}(\lambda) =  
\int_{\Xdn}\ F({f_m};\ [\lambda]x) g_n(x) d^n\sigma(x) 
\end{equation}
defined for $\lambda \in \bC_+$ and 
\begin{equation}
G_{(f_m,g_n)}'(\lambda) =  
\int_{\Xdn}\ F'({f_m};\ [\lambda]x) g_n(x) d^n\sigma(x) 
\end{equation}
defined for $\lambda \in \bC_-$.  
For an arbitrary function 
$g_n \in {\cal D}(\Xdn)$, these two holomorphic functions  
are distinct from each other.  
Now, if $g_n$ is taken in ${\cal D}({\cal U}^n_{h(x_0)})$,  
the r.h.s. of Eqs. (\ref{bvg2}) and (\ref{bvg'2}) coincide in view of
local commutativity, and therefore these two holomorphic functions
admit a common holomorphic extension $G_{(f_m,g_n)}(\lambda)$ in 
$\bC \setminus \Rpc$ whose boundary values on 
$\bR \setminus {0}$ satisfy the properties a) and b) of the theorem. 
(in view of Eqs. (\ref{bvg1})---(\ref{bvg'2})).   
\vskip 12pt

\noindent {\bf Proof of Lemma \ref{analcomp}} \HB
We concentrate on part a) of the lemma, part b) being  
completely similar.
\ifundefined{completionApp}{}\else 
At first, the fact that the function $(z,\ \lambda) \mapsto
H([\lambda]z)$ can be analytically continued in ${\cal Z}_{n\, +} \times \bC_+$ 
is a result of purely geometrical nature (based on the tube theorem)
which can be obtained as a direct application of  
lemma \ref{clemma} (ii) of Appendix \ref{compl}.  
In fact, for each point $x \in \JJ_n^{(r)}$,
the set $\{z = [\lambda]x;\ \lambda \in{\Bbb C}_+\}$ is contained
in $\Delta$ (namely in ${\cal Z}_{n\, +}$, as it directly follows from
Eq. (\ref{speclor}) and from the definitions of
$\JJ_n^{(r)}$ and ${\cal Z}_{n\, +}$). 
One can even check that each point $x \in \JJ_n^{(r)}$ is on the 
edge of a small open tuboid $\tau(x)$ contained in ${\cal Z}_{n\, +}$ 
such that $\{z = [\lambda]z';\ z' \in \tau(x),\ \lambda \in \bC_+\} 
\subset {{\cal Z}_{n\, +} \cup \VV} \subset \Delta$. 
On the other hand, for each point  
$z \in {\cal Z}_{n\, +}$ there exists a neighbourhood 
$\delta_+(z) $ of the real positive axis and a neighbourhood 
$\delta_-(z) $ 
of the real negative axis in the complex $\lambda$-plane, such that 
the   set
$\{[\lambda]z;  
\,\, \, \lambda \in \delta^+(z) \cup \delta^-(z)\}$  
is contained in $\Delta$: 
for $\lambda \in \delta_+(z)$ and 
$\lambda \in \delta_-(z) $
the corresponding subsets are respectively contained in ${\cal Z}_{n\, +}$  
and in ${\cal Z}_{n\, -}$. Therefore, the assumptions of  
lemma \ref{clemma} (ii) of Appendix \ref{compl}  
are fulfilled (by choosing the set $Q$ of the latter as a subset of
$\tau(x)$  and $D' = {\cal Z}_{n\, +}$ after an appropriate adaptation of the variables).  
In order to see that the new domain thus obtained (i.e.
$\{z = [\lambda]z';\ z' \in {\cal Z}_{n\, +},\ \lambda \in \bC_+\}$ 
yields an enlargement of $\Delta$, it is sufficient to notice that
every real point $x$ such that at least one component $x_j - x_{j-1}$ is 
time-like is transformed by any complex transformation $[\lambda]$ into
a point outside ${{\cal Z}}_{n\, \pm}$ and this is of course also true for all points 
$z \in {\cal Z}_{n\, +}$ tending to such real (boundary) points
(the neighbourhoods $\delta^{\pm}(z)$ becoming arbitrarily thin
in such limiting configurations).  
The second statement of the lemma precisely deals with these limiting
real configurations and with the fact that the analyticity  
of the boundary value $H([\lambda]x)$ in $\{\lambda \in \bC_+\}$ is maintained for 
all $x \in \Xdn$. The boundary value relations (\ref{BV1}) then follow from the fact 
that every point $x$ is a limit of points $z \in {\cal Z}_{n\, +}$ and that the latter are  
always such that $[\lambda]z \in {\cal Z}_{n\, +}$ for $\lambda > 0$ and   
$[\lambda]z \in {\cal Z}_{n\, -}$ for $\lambda < 0$.    
In order to avoid too subtle an argument for justifying the analyticity of the limit  
$H([\lambda]x)$ in $\{\lambda \in \bC_+\}$, 
we prefer to rely on an assumption of 
tempered growth (of the form (\ref{tempcond})) 
for $H$; the latter allows one to give an alternative version of
the analytic completion procedure 
which is based on the Cauchy integral representation,
and thereby includes the treatment of the boundary values. 
\fi
\vskip 12pt 

For $z = (z_1,\ldots,\ z_n) \in \bC^{nd}$,
we adopt the coordinates
\begin{equation}
\zeta_1 = z_1,\ \ \ \zeta_k = z_k - z_{k-1}\ {\rm for}\ 
1<k \le n,\ \ \ {\rm u}_j = \zeta_j^{(0)} + \zeta_j^{(d)},\ 
{\rm v}_j = \zeta_j^{(0)} - \zeta_j^{(d)},\ {\rm for}\ 1 \le j \le n ,
\label{zeta}
\end{equation}
\begin{equation}
r_j = (\zeta_j^{(1)},\ldots,\ \zeta_j^{(d-1)}).
\end{equation}
For every 
$z = (z_1,\ldots,\ z_n) \in {\cal Z}_{n\, +}$, we define
$G(z,\ \lambda) = H([\lambda]z)$.   
Easy computations using the  tempered growth condition show that
$G(z,\ \lambda)$ is a holomorphic function of 
$z$ and $\lambda = \rho \e^{i\theta}$ for $z \in {\cal Z}_{n\, +}$,
$\rho \in \Rpo$ and 
\begin{equation}
|\sin \theta | < {\kappa \over 2(1+2M)},\ \ 
\kappa = \min_j \left ( 1- {\|\Im r_j\|^2 \over \Im {\rm u}_j\,\Im {\rm v}_j}\right ),
\end{equation}
with
\begin{equation}
M = {1 \over \mu} \max_j\ \max \{ |\Re {\rm u}_j|,\ |\Re {\rm v}_j|\},\ \ 
\mu = \min_j\ \min \{ \Im {\rm u}_j,\ \Im {\rm v}_j\} ,
\end{equation}
which (for such values) satisfies bounds of the following form:  
\begin{equation}
|G(z,\ \lambda)| \le
K_1 (|\lambda| +1/|\lambda|)^L \left (
{1\over \mu\kappa} + \max_j |\zeta_j| \right )^L\ 
\le K_2 (|\lambda| +1/|\lambda|)^L \left (
{\rm dist}(z,\partial {\rm T}_n)^{-1}+ \max_j |\zeta_j| \right )^L ,
\label{bdintube}
\end{equation}
where $K_1$, $K_2$ are suitable constants. 

On the other hand if $z$ is real and 
$z \in \JJ_n^{(r)}$, i.e. 
${\rm u}_j >0$ and ${\rm v}_j <0$ for all $j$ (with the notations of Eq. (\ref{zeta})),
then $[\lambda] z \in {\cal Z}_{n\, +}$ whenever $\Im \lambda >0$ and
\begin{equation}
|H([\lambda] z)| \le 
K (|\lambda| +1/|\lambda|)^L \left (
{1\over \Im \lambda}\,\max_j\,(1/\Re {\rm u}_j - 1/\Re {\rm v}_j) +
\max_j |\zeta_j| \right )^L.
\label{bdonJ}
\end{equation}
This shows that $H([\lambda+i0] z)$ is a 
tempered distribution in 
$\lambda \in \bR$ with values in the polynomially bounded functions of $z$
on $\JJ_n^{(r)}$ (actually in the $\CC^\infty$ functions of $z$,
as the $z$ derivatives of $H$ and $G$ satisfy similar bounds).
When $\lambda <0$, as already noted, one has $[\lambda] z \in \VV$ and
$G(z,\ \lambda) = H([\lambda] z)$ is analytic in
$z$ and $\lambda$. Hence, for $z \in \JJ_n^{(r)}$, 
$G(z,\ \lambda+i0)$ is well-defined as a 
tempered distribution in 
$\lambda \in \bR$ with values in the polynomially bounded functions of $z$
on $\JJ_n^{(r)}$ and is the boundary value of a function holomorphic in $\bC_+$ and
bounded by the r.h.s. of Eq. (\ref{bdonJ}). 
For $\lambda \in \bC_+$ this function can be computed by the 
Cauchy formula:
\begin{equation}
G(z,\ \lambda) = 
{1\over 2\pi i} (i + \lambda -1/\lambda)^{2L+2}\,
\int_{\bR}
{G(z,\ \lambda'+i0) \over (i+ \lambda' - 1/\lambda')^{2L+2}
(\lambda'-\lambda)}d{\lambda'}.  
\label{Cau}
\end{equation}
As shown by Eq. (\ref{bdintube}), the r.h.s. of
this formula continues to make sense
for $z\in{\cal Z}_{n\, +}$ and defines a function of $z$ holomorphic and of
tempered growth in ${\cal Z}_{n\, +}$, with values in the functions of $\lambda$
holomorphic in $\bC_+$ and continuous on $\ovl{\bC_+}\setminus \{0\}$. 
Therefore it
has a boundary value in the sense of distributions as $z$ tends to the reals.
For real $\lambda \not= 0$, this boundary value coincides with 
$G(z,\ \lambda)$ (in the sense of distributions) when
$z \in \JJ_n^{(r)}$, hence (in view of the analytic continuation principle
extended by the edge-of-the-wedge theorem) 
the rhs of Eq. (\ref{Cau}) coincides with 
$G(z,\ \lambda)$ for all $z\in{\cal Z}_{n\, +}$ and all real $\lambda\not= 0$.
The formula (\ref{Cau}) thus holds for all $z \in {\cal Z}_{n\, +},\ \lambda \in \bC_+$ 
and, in the
sense of distributions, when $z$ tends to the reals;  
moreover, the relations (\ref{BV1}) hold in this limit
\ifundefined{completionApp}{}\else
as explained above in the geometrical analysis\fi.

The previous argument could be identically repeated for part b) of the lemma,  
replacing ${{\cal Z}}_{n\, \pm}$ by ${{\cal Z}'}_{n\, \pm}$ etc... and $\bC_+$ by $\bC_-$, 
since (as one can check directly) 
for each point $x \in {\JJ'}_n^{(r)}
 = -{\JJ'}_n^{(l)}$, the set $\{z = [\lambda]x;\ \lambda \in \bC_- \}$ 
is contained in ${{\cal Z}'}_{n\, +}$.  
\vskip 12pt

\noindent{\bf Remark. }  
Using the vector-valued analyticity provided by Lemma \ref{vecval} below,
it is possible to carry over the analysis of Bisognano and Wichmann without
change to the de Sitterian case. The above proof (also valid in the Minkowskian
case) aims at a clear distinction of the part of this theory which does not    
depend on positivity.

\subsection{Physical interpretation}

The following theorem gives a thermal 
physical interpretation to the weak 
spectral condition we have introduced.

\begin{thm} (KMS condition)\\
For every pair of bounded regions ${\cal O}_1$, ${\cal O}_2$ of 
${\cal U}_{h(x_0)}$,   the   correlation functions between 
elements of the corresponding
polynomial algebras  ${\cal P}({\cal O}_1)$, ${\cal P}({\cal O}_2)$ of a 
 field on $X_d$ satisfying the previous postulates  
enjoy a KMS condition with respect to the time-translation group 
$T_{h(x_0)}$ whose 
temperature  is  ${\rm T} =1/2\pi R$.
\label{kms}
\end{thm}

\noindent{\bf Proof.}\HB
Being given any general  correlation  function 
$(\Omega, \Rep({  f} )\Rep( {g} )\Omega)$
between arbitrary elements 
$\Rep(f ) \in {\cal P}({\cal O}_{1})$ and  
$\Rep(g ) \in {\cal P}({\cal O}_{2})$,
with $ f = (f_0, f_1,\ldots,f_m,\ldots),   
\ g = (g_0, g_1,\ldots,g_n,\ldots), 
(f_m \in {\cal D}({\cal O}_1^m),
g_n \in {\cal D}({\cal O}_2^n))$,   
we consider, for each ``time-translation'' $T_{h(x_0)}(t)$, the transformed quantities
\begin{equation}
{\cal W}_{(f,g)}(t) = (\Omega, \Rep({  f} )\Rep( {g_{\{e^{t/R}\}}} )\Omega) 
\label{correl} 
\end{equation}
and
\begin{equation}
{\cal W'}_{(f,g)}(t) = (\Omega,\Rep( {g_{\{e^{t/R}\}}} )\Rep({f})\Omega) 
\label{correl'} 
\end{equation}
(the notation $g_{\{e^{t/R}\}}$ being as in Eq. (\ref{Lambdaact}), with
${\Lambda_r} = [\lambda], \lambda = e^{\frac{t}{R}}$). 

In view of Theorem \ref{bw}, 
one can introduce the function 
$G_{(f,g)}(\lambda) = \sum_{m,n} G_{(f_m,g_n)}(\lambda)$, 
which is holomorphic for  
$ \lambda = e^{\frac{t}{R}} \in \bC\setminus \Rpc$ and admits  
continuous boundary values $G_{(f,g)}^\pm$
on $\Rpo$ from the upper and lower half-planes 
given respectively (in view of Eqs. (\ref{defG}) and (\ref{borch})) by:  
\begin{equation}
G_{(f,g)}^+(\lambda) = \sum_{m,n}  
\langle\WW_{m+n},\ f_m\otimes {g_n}_\lambda\rangle\ 
= (\Omega, \Rep({  f} )\Rep( {g_{\{e^{t/R}\}}} )\Omega), 
\end{equation}
\begin{equation}
G_{(f,g)}^-(\lambda) = \sum_{m,n}  
\langle\WW_{m+n},\ {g_n}_\lambda\otimes f_m\rangle\ 
= (\Omega,\Rep( {g_{\{e^{t/R}\}}} )\Rep({f})\Omega). 
\end{equation}
This readily implies that the function  
${\rm W}_{(f,g)}(t) = G_{(f,g)}(e^{\frac{t}{R}})$  
is holomorphic in the strip 
$0 < \Im t < 2\pi R$ 
and that it admits continuous boundary values on the edges of this strip  
which are: 
\begin{equation}
\lim_{\epsilon \to 0^+} 
{\rm W}_{(f,g)}(t + i\epsilon ) = 
{\cal W}_{(f,g)}(t), 
\ \ \ \ \lim_{\epsilon \to 0^+} 
{\rm W}_{(f,g)}(t + 2i{\pi}R  -i\epsilon)
= {\cal W'}_{(f,g)}(t).  
\label{kmsvec}
\end{equation}
The latter express the fact that all the field observables localized in
$ {\cal U}_{h(x_0)} $ and submitted to the time-translation group $ T_{h(x_0)}$ 
satisfy a KMS-condition at temperature ${\rm T} = (2\pi R)^{-1}$.  
\vskip 1cm

The previous property must be completed by
the following results: 
\vskip 12pt

\noindent
{\bf i) Periodicity in the complex time variable }

\noindent
Since $f$ and $g$ are localized respectively in ${\cal O}_1$ 
and ${\cal O}_2$, it follows from local commutativity that 
the function ${\rm W}_{(f,g)}(t)$ 
can be analytically continued across the part of the line
$\Im t = 0$ (and therefore $\Im t = 2n\pi R, \  n \in {\Bbb Z}$)
on which the two matrix elements of Eqs. (\ref{correl}) and (\ref{correl'}) 
are equal. One concludes that 
the function ${\rm W}_{(f,g)}(t)$ 
is holomorphic and periodic with period $2i\pi R$ in the 
following cut-plane  
${{\Bbb C}}^{cut}({\cal O}_1,{\cal O}_2)$  
which is {\em connected} (in particular) {\em if ${\cal O}_1$ and
${\cal O}_2$ are space-like separated}: 
\begin{equation}
{{\Bbb C}}^{cut}({\cal O}_1,{\cal O}_2)= 
\bigcap_{x_1 \in {\cal O}_1 ,x_2\in {\cal O}_2 } {{\Bbb C}}^{cut}_{x_1,x_2}, 
 \end{equation}
where
\begin{equation}
{{\Bbb C}}^{cut}_{x_1,x_2 }=\{t\in {{\Bbb C}};\;\Im t \not= 2n\pi R,\; n\in 
{\Bbb Z}\}
\cup \{t;\; t-2 i n\pi R\in I_{x_1,x_2}, \; n\in {\Bbb Z}\}.\label{ccut}
\end{equation}
and for any pair $(x_1, x_2)$  we have set 
\begin{equation}
I_{x_1,x_2} = \{t \in \bR: \ (x_1- [e^{-{\frac{t}{R}}}]{x_2})^2 < 0\}
\end{equation}
\vskip 1cm

\noindent
{\bf ii) The antipodal condition} 

The following property, relating by analytic continuation the field
observables localized in the region 
${\cal  U}_{h(x_0)}$ with those localized in the antipodal region 
\begin{equation}
{\check{\cal  U}}_{h(x_0)} = \{x\in X_d, -x\in   {\cal  U}_{h(x_0)}\}=
\{x = (x^{(0)},\vec{\x},x^{(d)}) \in X_d,\  \check{x} = (-x^{(0)},\vec{\x},-x^{(d)}) \in {\cal  U}_{h(x_0)}\}
\end{equation}
can also be obtained as a by-product of theorem \ref{bw}. 
\vskip 5pt

With each sequence  
$ g = (g_0, g_1,\ldots,g_n,\ldots)$ such that $ g_n \in {\cal D}({\cal U}_{h(x_0)}^n)$ 
let us associate the sequence
$ {\check{g}} = ( 
{\check{g}}_0,  
{\check{g}}_1, \ldots, 
{\check{g}}_n, \ldots)$,   
where $   
{\check{g}}_n(x_1,\ldots,x_n) = g_n^{\leftarrow}(\check{x_1},\ldots,\check{x_n}) 
= g_n(\check{x_n},\ldots,\check{x_1})$.    
Since (for each $n$) one has $ {\check{g}}_n \in 
{\cal D}({\check{\cal U}}_{h(x_0)}^n)$, it follows that $\Rep({\check{g}})$ belongs to
${\cal P}(
{\check{\cal  U}}_{h(x_0)})$. 

Let us also note that for the Lorentz transformation $[\lambda] = [-1]$, one has 
${g_n^{\leftarrow}}_{-1} = {\check{g}}_n$ and therefore, for all $\lambda > 0$,  
${g_n^{\leftarrow}}_{-\lambda} ={\check{g}_n}{}_{\lambda}$.  

We then see that the holomorphic function $ G_{(f,g)}(\lambda)$ introduced above
satisfies (in view of Eq. (\ref{Gantipod})) the following relations: 

for all $\lambda > 0$, 
\begin{equation}
G_{(f,g)}(-\lambda) = \sum_{m,n} 
\langle\WW_{m+n},\ f_m\otimes {g_n^{\leftarrow}}_{-\lambda}\rangle =  
\sum_{m,n} \langle\WW_{m+n},\ {g_n^{\leftarrow}}_{-\lambda}\otimes f_m\rangle. 
\end{equation}
and therefore in view of Eq. (\ref{borch}):  
\begin{equation}
G_{(f,g)}(-e^{t/R}) =  
(\Omega, \Rep({f} )\Rep({{\check{g}}_{\{e^{t/R}\}}} )\Omega). 
= (\Omega,\Rep({{\check{g}}_{\{e^{t/R}\}}} )\Rep({f})\Omega). 
\end{equation}

We can then state the following

\begin{prop}
(antipodal condition)\\
Being given arbitrary observables $\Rep({f})$ and $\Rep({g})$ in ${\cal P}({\cal U}_{h(x_0)})$ 
and the corresponding observable $\Rep({\check{g}})$ in 
${\cal P}({\check{\cal U}}_{h(x_0)})$,  
the following identities hold:

$\forall\  t \in {\Bbb R}$,  
\begin{equation}
{\rm W}_{(f,g)}(t + i\pi R) = (\Omega, \Rep({f} )\Rep({{\check g}_{\{e^{t/R}\}}} )\Omega) 
= (\Omega, \Rep({{\check g}_{\{e^{t/R}\}}})\Rep({f})\Omega) 
\label{antipod}
\end{equation}
\end{prop}
\vskip 12pt

\noindent
{\bf The geodesic and antipodal spectral conditions}
\vskip 1cm

We can introduce an ``energy operator'' 
${\cal E}_{h(x_0)}$ associated with the geodesic 
${h(x_0)}$ by considering in ${\cal H}$ the continuous unitary representation 
$\{ U^t_{h(x_0)};\linebreak[0] t\in {{\Bbb R}}\}$ of the 
time-translation group 
$T_{h(x_0)}$ and its spectral resolution 
\begin{equation}
U^t_{h(x_0)} = \int_{-\infty}^\infty e^{i\omega t} dE_{h(x_0)}(\omega);
\label{unitario}\end{equation}
This defines (on a certain dense domain of ${\cal H}$ containing
$\Rep(\BB)\Omega$) the self-adjoint operator
\begin{equation}
{\cal E}_{h(x_0)}=\int_{-\infty}^\infty \omega dE_{h(x_0)}(\omega).
\end{equation}
For any pair of   vector states   
$\Psi^{(1)}= \Rep({f}^\star )\Omega$,
$\Psi^{(2)}= \Rep({g})\Omega$, 
the corresponding
correlation function given in Eq. (\ref{correl}) 
can be written as follows:
\begin{equation}
{\cal W}_{(f,g)}(t) = 
( \Rep({f}^\star )\Omega, U^t_{h(x_0)}
\Rep(g)\Omega), 
\label{ruota}
\end{equation}
which shows that ${\cal W}_{(f,g)}(t)$ is a continuous and bounded function.
In view of Eq. (\ref{unitario}) it can be expressed as the Fourier transform 
of the bounded measure 
\begin{equation}
\wt{\cal W}_{(f,g)}(\omega)
= (\Rep({f}^\star)\Omega,dE_{h(x_0)}(\omega) \Rep({g})\Omega) 
\label{tutor}
\end{equation}
Similarly, one has:  
\begin{equation}
{\cal W'}_{(f,g)}(t) = 
( \Rep({g}^\star )\Omega, U^{-t}_{h(x_0)}
\Rep(f)\Omega), 
\label{ruota'}
\end{equation}
which is the Fourier transform of 
\begin{equation}
\wt{\cal W}'_{(f,g)}(\omega)
= ( \Rep({g}^\star)\Omega,dE_{h(x_0)}(-\omega) \Rep({f})\Omega).   
\label{dicendo}
\end{equation}
Eqs. (\ref{tutor}) and (\ref{dicendo}) are valid for arbitrary  
${f}$ and ${g} $ in ${\cal B}$. Now, if ${f} $ and ${g}$
have supports in    $ {\cal U}_{h(x_0)}$, the functions 
${\cal W}_{(f,g)}(t)$ and ${\cal W}'_{(f,g)}(t)$ satisfy the KMS relations 
(\ref{kmsvec}) and their Fourier transforms satisfy (as bounded measures) 
the following relation 
which is equivalent to Eq. (\ref{kmsvec}):
\begin{equation}
\wt{\cal W}'_{(f,g)}(\omega)= 
e^{-2\pi R \omega} \wt{\cal W}_{(f,g)}(\omega).
\label{byby}
\end{equation}

Moreover, if we rewrite the antipodal condition (\ref{antipod}) as follows
(with notations similar to those of Eqs. (\ref{correl}) and (\ref{correl'})):
\begin{equation}
{\rm W}_{(f,g)}(t + i\pi R) = {\cal W}_{(f,{\check{g}})}(t)
= {\cal W}_{({\check{g}},f)}(t)
\end{equation}  
we see that
the corresponding Fourier transforms satisfy the following equivalent relations: 

\begin{equation}
\wt{\cal W}_{(f,\check{g})}(\omega)= 
\wt{\cal W}_{(\check{g},f)}(\omega)= 
e^{-\pi R \omega} \wt{\cal W}_{(f,g)}(\omega).
\end{equation}

We have thus proved the

\begin{thm}   

i)\ For every pair of  states 
$\Psi^{(1)}= \Rep({  f}^\star )\Omega$, 
$\Psi^{(2)}= \Rep({  g} )\Omega$
in ${\cal P}({\cal  U}_{h(x_0)})\Omega $, 
the corresponding matrix elements 
of the spectral measure
$dE_{h(x_0)}(\omega)$ satisfy the following {\rm geodesic spectral condition}: 
\begin{equation}
 ( \Rep({g}^\star )\Omega,dE_{h(x_0)}(-\omega) \Rep({f})\Omega
)=e^{-2\pi R \omega}
 ( \Rep({  f}^\star )\Omega,dE_{h(x_0)}(\omega) \Rep({g})\Omega
)\label{tyty}
\end{equation}
ii)\ Moreover, the previous matrix elements of the spectral measure are also
related to a third one which involves the antipodal state 
$\Rep({\check{g}})\Omega$ 
in ${\cal P}({\check{\cal  U}}_{h(x_0)})\Omega $, 
by the following {\rm antipodal spectral condition}:
\begin{equation}
( \Rep({f}^\star )\Omega,dE_{h(x_0)}(\omega) \Rep({\check{g}})\Omega)  
= e^{-\pi R \omega}
( \Rep({f}^\star )\Omega,dE_{h(x_0)}(\omega) \Rep({g})\Omega). 
\label{tyty'}
\end{equation}
\label{pr4}
\end{thm}

\begin{remark}{\em     
i) The geodesic spectral condition (\ref{tyty}) 
gives a precise content to the statement that 
in the region ${\cal U}_{h(x_0)}$ corresponding to  an observer living 
on the geodesic ${h(x_0)}$, the  energy measurements (relative to this observer)
give exponentially damped expectation values in the range of negative energies. 
In the limit of flat space-time  the l.h.s. of Eq. (\ref{tyty}) would be  
equal to zero for $\omega >0$, which corresponds to recovering the usual 
spectral condition of ``positivity of the energy''.

ii) The antipodal spectral condition (\ref{tyty'}) 
asserts that the spectral measure 
$ dE_{h(x_0)}$ has exponentially damped matrix elements, 
in the high energy limit, between states 
localized in the mutually antipodal regions ${\cal  U}_{h(x_0)}$ and 
${\check{\cal  U}}_{h(x_0)}$.  
}\end{remark}

\begin{remark}{\em     
All the features that have been discussed in this section are also naturally interpreted  in terms of the existence of an antiunitary involution $J$ 
relating the algebras ${\cal P}({\cal U}_{h(x_0)})$ and  
${\cal P}({\check{\cal U}}_{h(x_0)})$ 
and the validity of the corresponding Bisognano-Wichmann duality theorem for the 
Von Neumann algebras ${\cal A}({\cal U}_{h(x_0)})$ and 
${\cal A}({\check{\cal  U}}_{h(x_0)})$   
\cite{[Ara],[BW]}}.
\end{remark}

\section{A consequence of positivity and weak spectral condition: the Reeh-Schlieder property}
In this section we wish to show that the vector-valued
distributions $f_n \mapsto \langle \Phi_n^{(b)},\ f_n  \rangle,$ 
(which are provided by the GNS construction, 
see Eq. (\ref{vecvaldis})), are boundary values of vector-valued 
functions holomorphic in the tuboids
$\ZZ_n = \ZZ_{n\,+} = {\rm Z}_{n,\ d+1} \cap {\Xcdn}$, where
 \begin{equation}
{\rm Z}_{n,\, d+1} = \left\{ z\in {\bC^{n(d+1)}};\  
y_1 \in V_+ ,\ y_j - y_{j-1} \in V_+ ,\ j = 2,\ldots,\  n  \right \} ,  
\end{equation}
with, in particular, the Reeh-Schlieder property as a consequence.
Let us  also recall the definition of 
$
{\ZZ'}_n = {\ZZ'}_{n\, +} =  
 \left\{ z\in {\Xcdn};\  
y_n \in V_- ,\ y_j - y_{j-1} \in V_+ ,\ j = 2,\ldots,\  n  \right \}.  
$
 
In the Minkowskian, flat, $d$-dimensional case, assuming the temperedness 
condition, as a consequence of the
spectral condition (see e.g. \cite{[J]}), 
the vector-valued distribution $\Phi_n^{(b)}$
is the Fourier transform of a vector-valued 
tempered distribution with support in the cone dual to the base of the tube
${\rm Z}_{n,\, d}$. Hence $\Phi_n^{(b)}$ is the boundary value of a 
function holomorphic in ${\rm Z}_{n,\, d}$. This fact can also be seen,
in this case, by using the maximum principle and 
the fact that the distinguished
boundary of  ${\rm Z}_{n,\, d}$ is $\bR^{dn}$.
These tools are not available in the de Sitterian case, but,
as mentioned before, a theorem of V.~Glaser, stated below, can be used
in conjunction with the positivity and weak spectral conditions, to prove:

\begin{thm}
\label{vecval}
There exists, for each $n \ge 1$, a function $\Phi_n$ holomorphic in
${\ZZ_n}$ with values in $\HH$ such that $\Phi_n^{(b)}$ is the boundary
value of $\Phi_n$ in the sense of 
distributions and of the
Hilbert space topology.
\end{thm} 
\vskip 12pt
 
\noindent
Theorem \ref{vecval} implies the Reeh-Schlieder property:

\begin{thm} 
\label{ree} (Reeh-Schlieder)
For every open subset  $\cal O$ of $X_d$, the vacuum is cyclic for the
algebra of all field polynomials localized in $\cal O$.
\end{thm}

\noindent {\bf Proof.}
For every $\Psi \in \HH$ and every $n \ge 1$, the distribution
$(\Psi,\ \Phi_n^{(b)})$ is the boundary value of the function
$z \mapsto (\Psi,\ \Phi_n(z))$, holomorphic in $\ZZ_n$. If 
$\cal O$ is an open subset of $X_d$ such that 
$(\Psi,\ \langle\Phi_n^{(b)},\ \vhi\rangle)$ vanishes for
every $\vhi \in \DD({\cal O}^n)$ then it vanishes for all $\vhi \in \DD(X^d)$
by analytic continuation, and since the vector space ${\cal P}(X_d)\Omega$ 
is dense in $\cal H$, this implies that $\Psi = 0$. Therefore
the vector space generated by
$\{\langle \Phi_n^{(b)},\ \vhi\rangle\ :\ 
\vhi \in \DD({\cal O}^n),\ n \in \bN\}$ is dense in $\HH$.
\vskip 12pt

\noindent
For proving 
theorem \ref{vecval} we shall make use of the following immediate consequence
of the weak spectral condition
\vskip 12pt

\begin{prop}
\label{prodanal} 
For each pair of integers $(m,n)$, the function 
$(w,\ z)\mapsto {\rm W}_{m+n}(w,\ z)$,
($ w \in \Xcdm$, $z \in \Xcdn$), is holomorphic in the 
corresponding topological product $ {\ZZ'}_m \times {\ZZ}_n $.  
\end{prop}

\noindent
We are now in a position to apply 
the following theorem proved by
V.~Glaser in \cite{[G1]} (see also a restatement in \cite{[G2]}).
We suppose given a finite sequence of non-empty
domains $U_n \subset \bC^{N_n}$, $1 \le n \le M$, where the $N_n$
are integers and $N_n \ge 1$. We  set $N_0 =0$, i.e. $U_0$ 
can be considered as consisting of a single point.
$U_n^*$ will denote the complex conjugate domain of $U_n$.
For $n \ge 1$, $\lambda_n$ denotes the Lebesgue measure in $\bC^{N_n} \equiv {\Bbb R}^{2 N_n}$  .
\vskip 12pt

\noindent
{\bf Glaser's theorem 1}
\it For each pair of integers $(n,\ m)$ with $0 \le n,\ m \le M$, let
$(p_n,\ q_m) \mapsto A_{n\,m}(p_n,\ q_m)$ be a holomorphic
function on $U_n\times U_m^*$. (In particular $A_{0\,0}$ is just
a complex number.)
Then the following properties are equivalent:\HB
(G.0) For each $n \in [1,\ M]$, there is an open neighborhood 
$V_n$ of 0 in $\bR^{N_n}$  and a point $p_n \in U_n$ such that 
$p_n+V_n \subset U_n$ and, for each
sequence $\{f_n\}_{0 \le n \le M}$, $f_0 \in \bC$, 
$f_n \in \DD(V_n)$ for $n >0$,
\begin{equation}
\sum_{0 \le n,\ m \le M}
\int_{\bR^{N_n}\times \bR^{N_m}}
A_{n\,m}(p_n+h_n,\ \bar p_m+k_m)\,\bar f_n(h_n)f_m(k_m)dh_n\,dk_m \ge 0\ 
\end{equation}
(with an obvious meaning when $n$ or $m$ is equal to 0).\HB
(G.1) For every sequence $\{g_n\}_{0 \le n \le M}$, $g_0 \in \bC$, 
$g_n \in \DD(U_n)$ for $n >0$,
\begin{equation}
\sum_{0 \le n,\ m \le M}
\int_{U_n\times U_m} A_{n\,m}(p_n,\ \bar q_m)\, \bar g_n(p_n)\,g_m(q_m)\, 
d\lambda_n(p_n) d\lambda_m(q_m)
\ge 0.
\end{equation}
(G'.1) For each $n \in [1,\ M]$, there is an open subset $\omega_n$
of $U_n$ such that for every sequence 
$\{g_n\}_{0 \le n \le M}$, $g_0 \in \bC$,
$g_n \in \DD(\omega_n)$ for $n >0$,
\begin{equation}
\sum_{0 \le n,\ m \le M}
\int_{\omega_n\times \omega_m} A_{n\,m}(p_n,\ \bar q_m)\, 
\bar g_n(p_n)\,g_m(q_m)\, 
d\lambda_n(p_n) d\lambda_m(q_m)
\ge 0.
\end{equation}
(G.2) There is a sequence $\{f_{\nu,\ 0}\}_{\nu \in \bN}\in \bC$ and,
for each $n \in [1,\ M]$, a sequence 
$\{f_{\nu,\ n}\}_{\nu \in \bN}$ of functions holomorphic in $U_n$,
such that 
\begin{equation}
A_{n\,m}(p_n,\ q_m) = \sum_{\nu \in \bN}
f_{\nu,\ n}(p_n)\,\ovl{f_{\nu,\ m}(\bar q_m)}
\end{equation}
holds in the sense of uniform convergence on every compact subset of
$U_n\times U_m^*$, again
with an obvious meaning when $n$ or $m$ is equal to 0.\HB
(G.3) For every sequence $\{p_n \in U_n\}_{1\le n \le M}$,
and every finite sequence $\{a(n)_\alpha\}$ of complex numbers,
\begin{equation}
Q_p(a,\ a) = \sum_{n,\ m} \sum_{\alpha,\ \beta}
{a(n)_\alpha \,\bar a(m)_\beta \over \alpha !\,\beta !}\,
\partial_{p_n}^\alpha \,\partial_{\bar p_m}^\beta\, 
A_{n\,m}(p,\ \bar p) \ge 0 .
\end{equation}
(G.4) There is a particular sequence $\{p_n \in U_n\}_{1\le n \le M}$
such that, for every finite sequence $\{a(n)_\alpha\}$ of complex numbers,
$Q_p(a,\ a) \ge 0$.
\vskip 12pt

\noindent
\rm The following striking theorem, also proved in \cite{[G1]} is mentioned here for
completeness although it is not used in the proof of theorem \ref{vecval}:
\vskip 12pt

\noindent
{\bf Glaser's theorem 2}
\it Let $U$ be a non-empty simply connected domain in $\bC^N$ (with $N \ge 1$), 
and $F$ a distribution over $U$, such that, for every finite sequence 
$\{a_\alpha\}$ of complex numbers indexed by $N$-multiindices, 
\begin{equation}
\sum_{\alpha,\ \beta} {a_\alpha \,\bar a_\beta \over \alpha !\,\beta !}\,
\partial^\alpha\, \bar \partial^\beta F  \ge 0
\end{equation}
(in the sense of distributions). Then there is a function 
$(p,\ q) \mapsto A(p,\ q)$, holomorphic on $U\times U^*$ 
and possessing the properties (G.1)-(G.3) of Glaser's theorem 1 
(in the case $M=1$),
such that $F$ coincides with $p \mapsto A(p,\ \bar p)$. 
\vskip 12pt

\noindent{\bf Remarks.}\HB
\rm
1) The statement of Glaser's theorem 1  does not literally coincide with
the original in \cite{[G1]}, but it follows from the proofs given there.\HB
2) In the condition (G.1) one can equivalently require the $g_n$ to be
arbitrary complex measures with compact support contained in $U_n$.
Since any measure can be weakly approximated by finite linear combinations
of Dirac measures, the condition (G.1) is equivalent to

\noindent\sl
(G''.1) For every finite sequence 
$\{(c_{n,l},\ t_{n,l})\ :\ c_{n,l} \in \bC,\ 
t_{n,l} \in U_n,\ 0 \le n \le M,\ 1 \le l \le L\}$,
\begin{equation}
\sum_{n,\ m = 0}^M\,
\sum_{l,\ k = 1}^L\, 
c_{n,l} \ovl{c_{m,k}} A_{n\ m}(t_{n,l},\ \ovl{t_{m,k}}) \ge 0.
\label{points}\end{equation}
\rm

\noindent
3) Apart from condition (G.0), the properties mentioned in these
theorems are essentially invariant under 
holomorphic self-conjugated coordinate
changes and in fact \it the various $U_n$ can be replaced by connected 
complex manifolds which are separable at infinity (i.e. are unions of 
increasing sequences of compacts) \rm 
\ifundefined{GlaserApp}{}\else as it can be seen from the sketch
of the proof given in Appendix \ref{glproof}\fi.
\vskip 12pt

\noindent {\bf Proof of Theorem \ref{vecval}}\HB
Taking into account the previous remark 3),  
we shall apply Glaser's theorem 1 to the case when each $U_n $ is 
the domain $ \ZZ_n$ of the corresponding manifold $\Xcdn$ and
\begin{equation}
A_{0\,0} = 1,\ \ \ \ 
A_{n\,m}(z,\ w) = 
{\rm W}_{m+n}(w_\leftarrow,\  z) = 
{\rm W}_{m+n}(w_m,\ldots,w_1,\ z_1,\ldots,z_n,)
\label{GLW}
\end{equation}
with $n,\ m \in [0,\ M]$, $M$ being any fixed integer. 

In fact, in view of proposition \ref{prodanal} and of the remark that 
\begin{equation}
\ZZ_m^* = \{w = (w_1,\ldots,\ w_m) \in X^{(c)m}_d;\ w_\leftarrow =
(w_m,\ldots,\ w_1) \in {\ZZ'}_m \}, 
\end{equation}
it follows that for all pairs of integers
$(n,m)$ the functions defined by Eq. (\ref{GLW}) are holomorphic in the corresponding
domains $\ZZ_n \times \ZZ_m^*$.  
Now our aim is to prove that, as a
consequence of the positivity property (\ref{posit}), these functions
possess the properties (G.0)-(G.4) of Glaser's theorem 1. 
Let $a$ be a particular point of $X_d$ (e.g. $a = (0,\ldots,0,\ R)$).
It is clear that we
can define, for each $n \ge 1$, 
a holomorphic diffeomorphism $\sigma_n$ of an open  ball centered at 0 in
$\bC^{nd}$ onto a complex neighborhood $\NN_n$ of $a_n = (a,\ a,\ldots,\ a)$ in
$X_d^{(c)\,n}$ with the following properties:

\begin{enumerate}
\item $\sigma_n$ is self-conjugate, i.e. 
$\sigma_n(\bar z) = \ovl{\sigma_n(z)}$ for all $z$.

\item $\sigma_n(0) = a_n$

\item $\sigma_n$ maps the ``local tube''
\begin{equation}
\{z = x+iy \in \bC^{nd}\ :\ |z_j| < 1,\ 0< y_j,\ 1 \le j \le nd\}
\end{equation}
into $\ZZ_n \cap \NN_n$.
\end{enumerate}

\noindent
In $\sigma_n^{-1}(\NN_n\cap\ZZ_n)\times \sigma_m^{-1}(\NN_m\cap\ZZ_m^*)$,
there holds (in view of the distribution character of the boundary values
of the ${\cal A}_{n\ m}$ on $X_d^{n+m}$): 
\begin{equation}
|A_{n\,m}(\sigma_n(z),\ \sigma_m(z'))| \le 
K(\sum_j |\Im z_j|^{-r} + \sum_j |\Im z'_j|^{-r}),
\end{equation}
where $K>0$ and $r\ge 0$ may be taken independent of $n,\ m \in [1,\ M]$.

By composing $\sigma_n$ with $z_j = {\rm th}\,(\zeta_j/2)$, we obtain a
self-conjugate holomorphic diffeomorphism $\tau_n$ of 
the tube 
\begin{equation}
\{\zeta = \xi + i\eta \in \bC^{nd}\ :\ |\eta_j| < \pi/2,\ 1 \le j \le nd\}
\end{equation}
onto a complex neighborhood of $a_n$ in $X_d^{(c)\,n}$ such that
$\tau_n(0) = a_n$ and the image of the tube
\begin{equation}
\Theta_n = 
\{\zeta = \xi + i\eta \in \bC^{nd}\ :\ 0<\eta_j< \pi/2,\ 1 \le j \le nd\}.
\end{equation}
is contained in $\ZZ_n$. 
Let
\begin{equation}
B_{n\,m}(\zeta,\ \zeta') = 
A_{n\,m}(\tau_n(\zeta),\ \tau_m(\zeta')).
\end{equation}
The functions $B_{n\,m}$ are holomorphic in $\Theta_n \times \Theta_m^*$.
Since for $\zeta = \xi + i\eta\in \bC$,
\begin{equation}
\th(\zeta/2) = {\sh \xi +i\sin \eta \over
2|\ch(\zeta/2)|^2},
\end{equation}
the $B_{n\,m}$ satisfy
\begin{equation}
\begin{array}{l}
\displaystyle
|B_{n\,m}(\zeta,\ \zeta')| \le
K'\sum_j \left( {\e^{|\xi_j|} \over |\sin \eta_j|} \right )^r 
+K'\sum_j \left( {\e^{|\xi'_j|} \over |\sin \eta'_j|} \right )^r ,\\
\displaystyle \hbox to 4 truecm{\hfill}
\forall \zeta = \xi +i\eta \in \Theta_n,\ 
\zeta' = \xi' +i\eta' \in \Theta_m^*.\\
\end{array}
\end{equation}
They have boundary values $B_{n\,m}^{(v)}$ in the sense of generalized
functions over test-functions of faster than exponential decrease.
These boundary values satisfy, for each finite sequence 
$\{f_n\}$, $f_0 \in \bC$, $f_n \in \DD(\bR^{nd})$ for $n \ge 1$,
\begin{equation}
\sum_{n,\ m} \int B_{n\,m}^{(v)}(\xi,\ \xi')
f_n(\xi)\,\ovl{f_m(\xi')}\,d\xi\,d\xi'   \ge 0.
\label{Bpos}\end{equation}
Let now
\begin{equation}
\rho_{n,\ \veps}(\xi) = 
C(\veps)\,\exp \left( -\sum_{j=1}^{nd} (\xi_j^2/\veps)\right ) ,
\end{equation}
where $C(\veps)$ is chosen so that $\int \rho_{n,\ \veps}(\xi)\,d\xi =1$.
For each $\mu\in\bC^{nd}$, the function 
$\xi \mapsto \rho_{n,\ \veps}(\xi+\mu)$ is of gaussian decrease,
and depends holomorphically on $\mu$. In particular if $\mu_n \in \Theta_n$,
$\mu'_m \in \Theta_m^*$,
\begin{equation}
\begin{array}{l}
\displaystyle
\int_{\bR^{nd}\times \bR^{md}} 
B_{n\,m}^{(v)} (t,\ t')\, f_n(\xi)\,\rho_{n,\ \veps}(t-\mu_n -\xi)\,
\ovl{f_m(\xi')\rho_{m,\ \veps}(t'-\bar\mu'_m -\xi')}\,dt\,dt'\,d\xi\,d\xi'\\
\displaystyle =
\int_{\bR^{nd}\times \bR^{md}}
B_{n\,m}(t+\mu_n,\ t'+\mu'_m)\,f_n(\xi)\,\rho_{n,\ \veps}(t -\xi)\,
\ovl{f_m(\xi')\rho_{m,\ \veps}(t'-\xi')}\,dt\,dt'\,d\xi\,d\xi' ,\\
\end{array}
\end{equation}
since both sides define analytic functions in 
$\Theta_n\times \Theta_m^*$ whose boundary values for real $\mu_n$, $\mu'_m$
coincide.
The lhs satisfies the positivity conditions, by virtue of Eq. (\ref{Bpos}),
if we chose $\mu'_n = \bar \mu_n$ for all $n$.
It follows, by letting $\veps$ tend to 0 in the rhs,
that the functions $B_{n\,m}$ have the property (G.0) of Glaser's theorem 1 and
therefore all the properties (G.0)-(G.4) in the sequence of domains
$\{\Theta_n\}$. Coming back to the original variables, Glaser's theorem 1 
now shows that the same properties, in particular (G.2),
extend to the entire tuboid $\{\ZZ_n\}$.
We have thus proved the following

\begin{prop} 
\label{expan}
For any integer $M \ge 1$,
there exist a sequence $\{F_{\nu,\ 0}\in \bC\}_{\nu \in \bN}$ and,
for each integer $n \in [1,\ M]$, a sequence $\{F_{\nu,\ n}\}_{\nu \in \bN}$ 
of functions holomorphic in $\ZZ_n$, such that, for every 
$n$ and $m$ in $[1,\ M]$, $z \in \ZZ_n$, $w \in \ZZ_m^*$,
\begin{equation}
{\rm W}_{m+n}(w_\leftarrow,\  z) =
\sum_{\nu \in \bN}\, \ovl{F_{\nu,\ m}(\bar w)}\,F_{\nu,\ n}(z) ,
\label{2}\end{equation}
where the convergence is uniform on every compact subset of
$\ZZ_m^*\times\ZZ_n$.
\end{prop}

In particular
\begin{equation}
\label{schwarz}
{\rm W}_{2n}(\bar z_\leftarrow,\  z) =
\sum_{\nu \in \bN}\,
|F_{\nu,\ n}(z)|^2 ,
\end{equation}
(so that if the temperedness condition holds,
each $F_{\nu,\ n}$ has polynomial behavior at infinity and 
near the reals).

Let now $\{\vhi_m\}_{1\le m \le M}$ be a sequence
of test-functions, $\vhi_m \in \DD(X_d^m)$, $\vhi_0 \in \bC$. We 
continue to denote $\vhi_m$ a $\CC^\infty$ extension of $\vhi_m$ 
with compact support over $X_d^{(c)\,m}$. Let $\CC(m,\ \veps)$ be, for each 
$m \in [1,\ M]$ and $\veps \ge 0$, an $(md)$-cycle, contained in $\ZZ_m$
for $\veps >0$, equal to $X_d^m$ for $\veps = 0$, and continuously
depending on $\veps$. Using proposition
\ref{expan} and Schwarz's inequality,
we find, for any $z \in \ZZ_n$ ($n$ being fixed and $M$ chosen arbitrarily
such that $n \le M$),

\begin{equation}
\begin{array}{l}
\displaystyle
\left |\sum_{0 \le m \le M}
\int_{\CC(m,\ \veps)} \ovl{\vhi_m(w)}\,
{\rm W}_{m+n}(\bar w_\leftarrow,\ z)\,
d\bar w_1\wedge\ldots\wedge d\bar w_m \right|^2\\
= \left |\sum_{\nu \in \bN}\ [\sum_{0 \le m \le M}
\int_{\CC(m,\ \veps)} \ovl{\vhi_m(w)}\,
\ovl{F_{\nu,\ m}(w)}
d\bar w_1\wedge\ldots\wedge d\bar w_m] 
F_{\nu,\ n}(z)
\right|^2\\ 
\displaystyle\le\  \sum_{\nu \in \bN} 
|F_{\nu,\ n}(z)|^2 \ \times \\
\displaystyle
\sum_{0 \le m,\ k \le M}
\int_{\CC(m,\ \veps)\times \CC(k,\ \veps)}
\ovl{\vhi_m(w)}\,\vhi_k(w')\,
{\rm W}_{m+k}(\bar w_\leftarrow,\ w')\,
d\bar w_1\wedge\ldots\wedge d\bar w_m\wedge
dw'_1\wedge\ldots\wedge dw'_k .\\
\end{array}
\end{equation}

Taking Eq. (\ref{schwarz}) into account and
letting $\veps$ tend to 0 then yield: 
\begin{equation}
\begin{array}{l}
\displaystyle
\left |\sum_{0 \le m \le M}
\int_{X_d^m} \ovl{\vhi_m(w)}\,
{\rm W}_{m+n}(w_\leftarrow,\ z)\,
dw_1\ldots dw_m \right|^2\\
\displaystyle
\hbox to 4cm{\hfill}\le {\rm W}_{2n}(\bar z_\leftarrow,\ z)\ 
\left \Vert 
\sum_{0 \le m \le M} \int_{X_d} \Phi_m^{(b)}(w)\,\vhi_m(w)\, dw 
\right \Vert^2
.\\
\end{array}
\end{equation}
Since the latter holds for any (arbitrarily large) value of $M$,  
namely for a dense set of vectors ${\bf \Phi}(\vhi)\ \Omega$ in $\cal H$,
this shows that for every $n$ there is a vector 
$\Phi_n(z) \in \HH$ such that 
\begin{equation}
\left (\int_{X_d} \Phi_m^{(b)}(w)\,\vhi_m(w)\, dw,\ 
\Phi_n(z)\right) =
\int_{X_d} {\rm W}_{m+n}(w_\leftarrow,\ z) \ovl{\vhi_m(w)}\,dw.
\end{equation}

Integrating similarly in $z$ over a cycle such as $\CC(n,\ \veps)$,
and letting $\veps$ tend to 0 show that $\Phi_n$ admits  
$\Phi_n^{(b)}$ as its boundary value in the sense of distributions and
theorem \ref{vecval} follows.
\vskip 12pt

\noindent
{\bf Remarks}\HB
1. This proof is valid for some other spaces besides de Sitter space.
What is really used is that the space is real-analytic and that
the Wightman distributions are boundary values of functions 
${\rm W}_{m+n}(w_\leftarrow,\ z)$ holomorphic in products of the
form $U_m^*\times U_n$, where the $U_n$ are connected complex tuboids.\HB
2. Neither temperedness nor locality have been used. \HB
3. By using the PCT property, the BW analyticity and the Reeh-Schlieder property 
it is possible to restate the full Bisognano-Wichmann theorem in the de Sitter case. 
We do not give here the details.

\pagebreak 
\appendix
\ifundefined{completionApp}{}\else
\section{Appendix. A lemma of analytic completion}
\label{compl}
In this appendix we prove a simple lemma of analytic completion 
by applying the convex tube theorem, according to which
any function which is holomorphic in a tube
${\Bbb R}^n + i B $,  where $B $ is a domain in ${\Bbb R}^n$, 
can be analytically continued in the convex hull of this tube.
(See \cite{[BoM], [W], [E2], [E1]}).
$\bC_+$ denotes the upper half-plane.
\begin{lemma}\label{clemma}
(i) Let 
\begin{equation}
P =  
\{z \in \bC^N\ :\ |z_j| <1,\ \Im z_j > 0,\ \forall j=1,\ldots,\ N\}
\end{equation}
Let $D$ be a domain in $\bC^N$, containing $P$,
and $\Omega$ a domain in 
$\bC \times \bC^N$ of the form 
\begin{equation}
\Omega = \NN\cap (\bC_+ \times D),
\end{equation}
where $\NN$ is an open neighborhood, in $\bC^{1+N}$, of the set
\begin{equation}
\bigl((\bR \setminus \{0\})\times D\bigr) \cup
\bigl((\ovl {\bC_+} \setminus \{0\}) \times
\{z \in \bC^N\ :\ |z_j| <1,\ \Im z_j = 0,\ \forall j=1,\ldots,\ N\}\bigr).
\end{equation}
Then any function
holomorphic in $\Omega$ has a holomorphic extension in 
$\bC_+ \times D$.\HB
(ii) Let $D'$ be a domain in $\bC^N$, containing
\begin{equation}
Q = \{z \in \bC^N\ :\ |z_j| <1,\ \forall j=1,\ldots,\ N\},
\end{equation}
and $\Omega'$ a domain in 
$\bC \times \bC^N$ of the form 
\begin{equation}
\Omega' = (\bC_+ \times Q) \cup (\NN'\cap (\bC_+ \times D')),
\end{equation}
where $\NN'$ is an open neighborhood, in $\bC^{1+N}$, of
$(\bR \setminus \{0\})\times D'$.
Then any function
holomorphic in $\Omega'$ has a holomorphic extension in 
$\bC_+ \times D'$.
\end{lemma}

\begin{remark}\label{strip}

By setting $w = \e^{\pi\sigma}$ the upper half-plane can be replaced
by the strip $\{\sigma\ :\ 0<\Im \sigma <1\}$, and $\bR\ \setminus \{0\}$
by the boundary of that strip. 
\end{remark}
\noindent 1.
We start by proving Lemma \ref{clemma} (i) for the case when 
$D=P$. 
This follows from:

\begin{lemma}
\label{complaux}
Let $a \in (0,\ 1)$ and $\Delta'_a$ a domain in $\bC \times \bC^N$ of the form 
$\VV \cap (\bC_+\times P)$, where $\VV$ is an open
neighborhood in $\bC^{1+N}$ of
\begin{equation}
\begin{array}{l}
\{(w,\ z)\in \bC^{1+N}\ :\ w\in \bR\ :\ a< |w| <1/a\},\ \ z\in P\}\ \  \cup
\hbox to 2 truecm{\hfill}\\
\hbox to 0.5 truecm{\hfill}
\{(w,\ z)\in \bC^{1+N}\ :\ w\in \bC_+\cup(-1/a,\ -a)\cup(a,\ 1/a),\ \ 
|z_j| <1,\ \Im z_j = 0,\ \forall j=1,\ldots,\ N\}.\\
\end{array}
\end{equation}
Then any function $f$
holomorphic in $\Delta'_a$ has a holomorphic extension in the domain
\begin{equation}
\Delta_a = \bigcup_{0< \theta < \pi }
W_a(\theta) \times Z(\theta) ,
\label{holext}
\end{equation}
where:
\begin{equation}
Z(\theta) = \{z \in \bC^N\ :\ 
\forall j = 1,...\ N,\ \ \Im z_j > 0  ,\
2\,\Im \log \left ({1 + z_j \over 1-z_j}\right ) < \theta\ \} ,
\end{equation}
\begin{equation}
W_a(\theta) =
\{w \in \bC\ :\ 0< \Im w,\ 0 < \Im \Phi(w,\ a) < \pi - \theta \ \} ,
\end{equation}
\begin{equation}
\Phi(w,\ a) =
i\pi - \log \left ({w-a^{-1} \over w-a} \right ) \ -\
\log \left ({w+a \over w+a^{-1}} \right ),\ \ \ (\Im w \not= 0).
\end{equation}
\end{lemma}

\begin{remark}\em
The function $w \mapsto \Im\Phi(w,\ a)$ is the bounded harmonic function
in the upper half-plane with boundary values equal to 0 on the real
segments $(-a^{-1},\ -a)$ and $(a,\ a^{-1})$, and to $\pi$ on the other
real points. $\pi - \Im\Phi(w,\ a)$ is the sum of the angles under which
these two segments are seen from the point $w$.
\end{remark}

\noindent{\bf Proof.}
We shall make use (at several places and in several complex variables) of the 
following conformal map. 
For $A > 0$ and $B >0$,
we denote $L(A,\ B)$ the open lunule in the $ W $-plane bounded by
the real segment $[-A,\ A]$ and the circular arc going through the points
$-A$, $iB$, and $A$. This domain is conformally mapped onto the strip
$\{\lambda \in \bC\ :\ 0<\Im \lambda < 2 {\rm Arctg}\,(B/A)\}$
by the map
\begin{equation}
\label{confmap}
W \mapsto \lambda = \log\,
\left ( {A+W \over A-W} \right ) ,
\end{equation}
whose inverse is
\begin{equation}
\lambda \mapsto W = A\,\th(\lambda/2).
\end{equation}
Both the hypotheses of Lemma \ref{complaux} and the function $\Phi$ are left
invariant by the transformation $w \mapsto -1/w$. In fact, denoting
$b = a^{-1}$ and
\begin{equation}
\mu(w) = w - 1/w
\label{mu(w)}
\end{equation}
we have
\begin{equation}
{(w-b)(w+a) \over (w-a)(w+b)} \ \ =\ \ 
{\mu - (b-a) \over \mu +(b-a)} \ \ \equiv -1/\vhi(\mu) ,
\label{phi}
\end{equation}
\begin{equation}
\Phi(w,\ a) = \log\,\vhi(\mu).
\end{equation}
A function $f$ holomorphic in $\Delta'_a$ can be rewritten in the form:
\begin{equation}
f(w,\ z) = f_s(w,\ z)\ +\ (w+w^{-1}) f_a(w,\ z),
\end{equation}
where
\begin{equation}
f_s(w,\ z) = {1\over 2}(f(w,\ z) + f(-w^{-1},\ z)),\ \ \
f_a(w,\ z) = {1\over 2(w+w^{-1})}(f(w,\ z) - f(-w^{-1},\ z)).
\end{equation}
Both $f_s$ and $f_a$ are easily seen to have the properties postulated
for $f$ itself, and they are moreover invariant under the transformation
$w \mapsto -w^{-1}$. They can therefore be written as
holomorphic functions $F_{s,a}(\mu,\ z)$ of $\mu=w-w^{-1}$ and $z$.
We now perform the change of coordinates
\begin{equation}
\mu \mapsto \omega = \log \vhi(\mu),\ \ \
z_j \mapsto \zeta_j = 2\,\log\,((1 +z_j)/(1 -z_j)),
\label{zchange}
\end{equation}
i.e. define
\begin{equation}
G_{s,a}(\omega,\ \zeta) = F_{s,a}(\mu,\ z)\ \ {\rm with}\ \
\end{equation}
\begin{equation}
\mu = (b-a) {\rm th}\,(\omega/2),
\label{invmuchange}
\end{equation}
\begin{equation}
z_j = {\rm th}\,(\zeta_j/4).
\label{invzchange}
\end{equation}
The functions $G_{s,a}$ are holomorphic in a domain of the following form,
which is the image of the domain $\Delta'_a$ into the space of variables $(\omega,\zeta)$
(after taking the successive maps given in Eqs. (\ref{mu(w)}),(\ref{phi}) and (\ref{zchange}) into account):
\begin{equation}
\label{domain1}
U_1 = \VV_1 \cap 
\{(\omega,\ \zeta) \in \bC^{1+N}\ :
0< \Im \omega <\pi,\ \ 
0< \Im \zeta_j < \pi\} ,
\end{equation}
where $\VV_1$ is an open neighborhood, in $\bC^{1+N}$,
of the set
\begin{equation}
\begin{array}{r}
S_1 = \{(\omega,\ \zeta) \in \bC^{1+N}\ :\ 
\omega \in \bR,\ \ 0< \Im \zeta_j < \pi\}
\cup
\{(\omega,\ \zeta) \in \bC^{1+N}\ :\ 
0 \le \Im \omega < \pi,\ \ \Im \zeta_j = 0\}\\
= \{(\omega,\ \zeta) \in \bC^{1+N}\ :\ 
\omega \in \bR,\ \ 0\le \Im \zeta_j < \pi\} \cup
\{(\omega,\ \zeta) \in \bC^{1+N}\ :\ 
0 \le \Im \omega < \pi,\ \ \Im \zeta_j = 0\}.
\end{array}
\end{equation}
The domain $U_1$ of Eq. (\ref{domain1}) is not a tube. 
We can however inscribe in it increasing unions of topological products of 
lunules which are isomorphic to tubes.
In fact, for every $A > {1 \over \pi}$, there exists an $\veps > 0$ such that
$U_1$ contains
\begin{equation}
\begin{array}{r}
V_A = \{(\omega,\ \zeta) \in \bC^{1+N}\ :\ 
\omega \in L(A,\ \veps),\ \ \zeta_j \in L(A,\ \pi - 1/A)\}\\
\cup\  
\{(\omega,\ \zeta) \in \bC^{1+N}\ :\ 
\omega \in L(A,\ \pi - 1/A),\ \ \zeta_j \in L(A,\ \veps)\}.
\end{array}
\end{equation}
Using the conformal map (\ref{confmap}) in all variables, we can map $V_A$
into a tube whose holomorphy envelope is its convex hull. Returning
to the variables $(\omega,\ \zeta)$, and taking the limit 
$A \to  \infty$ shows that the functions $G_{s,a}$ are holomorphic in
the interior of the convex hull of $S_1$, namely
 \begin{equation}
\bigcup_{0< \theta < \pi }
\{(\omega,\ \zeta) \in \bC^{1+N}\ :\ 0< \Im \omega < \pi - \theta,\ \ 
0 < \Im \zeta_j < \theta,\  \forall j \}.
\end{equation}
This set is the image of the domain $\Delta_a$ 
introduced in Eq. (\ref{holext})
under the mapping
$w \mapsto \mu = w-w^{-1}\mapsto \omega$, $z \mapsto \zeta$ defined in 
Eq. (\ref{zchange}),
and therefore the assertion of
Lemma \ref{complaux} follows.
\vskip 5pt

\noindent
Lemma \ref{clemma} (i) in the special case $ D = P$ follows from the latter
by letting $ a $ tend to $ 0 $.

\noindent 2.
We now prove Lemma \ref{clemma} (ii) in the case when $D' = \rho\, Q$
for some real $\rho > 1$.
The proof of this is the same as that of Lemma \ref{complaux}, 
except that the change of coordinates (\ref{invzchange}) is replaced by
\begin{equation}
z_j = \exp(i\zeta_j), \ \ \ (1 \le j \le N).
\end{equation}
This again allows the use of the tube theorem.

\noindent 3.
Lemma \ref{clemma} (ii) follows from this by using chains of polydisks,
and (i) follows in the same way from the special case $D=P$ and (ii).

\fi

\ifundefined{BHWextApp}{}\else
\section{Appendix. A lemma of Hall and Wightman}
\label{BHW}
In \cite{[HW]}, Hall and Wightman prove the following lemma
\begin{lemma}
\label{hw}
Let $M\in L_+(\bC)$ be such that 
${\rm T}_+ \cap M^{-1}\,{\rm T}_+ \not= \emptyset$. 
There exists a continuous path 
$t \mapsto M(t)$ from the interval $[0,\ 1]$ into $ L_+(\bC)$ such
that $M(0) =1$, $M(1) = M$ and that, for every 
$z \in {\rm T}_+ \cap M^{-1}\,{\rm T}_+ \subset \bC^{d+1}$,
$M(t)z \in {\rm T}_+$ holds for all $t\in [0,\ 1]$.
\end{lemma}

This lemma is proved in \cite{[HW]} for the case $d+1 \le 4$
(a very clear exposition also appears in \cite{[SW]}).
It is extended to all dimensions in \cite{[J1]}.
We give another proof based on holomorphic continuation.
As noted in the above references,
if $M\in L_+(\bC)$ is such that the statement in Lemma \ref{hw}
holds, then it holds for $\Lambda_1 M\Lambda_2$ for any 
$\Lambda_1$, $\Lambda_2 \in L_+^\uparrow$, as well as for $M^{-1}$. 
It is therefore sufficient
to consider the case when $M$ is one of the normal forms classified
by Jost in \cite{[J1]}. $M$ can then be written in
the form:
\begin{equation}
M = \left (
\begin{array}{cc}
M_1(i) & 0\\
0 & M_2(i)
\end{array}
\right)
\end{equation}
where $t \mapsto M_1(t)$ is a one-parameter subgroup of the 
$p\times p$ Lorentz group, real for real $t$, with $p \le 3$, and
$t \mapsto M_2(t)$ is a one-parameter subgroup of the 
$(d+1-p)\times (d+1-p)$ orthogonal group, real for real $t$.
In the generic case $p \le 2$, $M_1(t) = 1$ if $p = 1$ and,
if $p=2$, $M_1(t) = [\exp at]$ for some real $a$ with $|a|\le \pi$.
We focus on this case
first. Replacing $M$ by $M^{-1}$ if necessary, we may assume $0<a\le \pi$. 
For any $z \in {\rm T}_+$ the set 
$\Delta(z,\ M) = \{t\in\bC\ :\ M(t)z \in {\rm T}_+\}$ is invariant under
real translations, i.e. is a union of open strips parallel to the
real axis. Let 
$$
E(M) = \{{\rm T}_+\cap M^{-1}{\rm T}_+\} =
\{z \in \bC^{d+1}\ :\ \bR\cup(i+\bR) \subset \Delta(z,\ M)\}.
$$
Denote $z(s) = (z^{(0)},\ z^{(1)},\ sz^{(2)},\ \ldots,\ sz^{(d)})$.
If $z \in E(M)$, then $z(s)\in E(M)$ for all $s \in [0,\ 1]$.
The set $\Delta(z(0),\ M)$ contains the segment $i[0,\ 1]$, and hence
$i[0,\ 1] \subset \Delta(z',\ M)$ for all $z'$ in a sufficiently small
neighborhood $\NN$ of $z(0)$. For $n\in \ovl{V_+}\setminus\{0\}$ and 
$b\in \bC$, with $\Im b \ge 0$, the function 
$(t,\ z) \mapsto H_{n,\ b}(t,\ z) = (n\cdot M(t)z+ b)^{-1}$
is holomorphic in 
$\{(t,\ z)\ :\ z \in E(M),\ t \in \Delta(z,\ M)\}$.
Applying Lemma \ref{clemma}ii), with $w$ replaced by
the variable $\sigma$ of Remark \ref{strip}, it
follows that $H_{n,\ b}$ is holomorphic
in $\{t\ :\ 0 < \Im t <1\}\times E(M)$. 
Let us now assume that for some $z \in E(M)$ and some $t \in i[0,\ 1]$ 
the corresponding point $\zeta = M(t)z$ belongs to the complement of ${\rm T}_+$;
then, as explained below, one can determine $n$ and $b$ satisfying the 
previous conditions and 
such that $n\cdot \zeta + b = 0$,      
which therefore contradicts the previously proved
analyticity property of the corresponding function $H_{n,\ b}$.
In fact,
for any complex point $\zeta = \xi + i\eta$ in the complement of 
${\rm T}_+$ (i.e. $\eta \notin V_+$), one can find 
$n \in \bR^{d+1}$ 
and $c \in \bR$ such that $n\cdot \eta +c = 0$, while 
$n\cdot r +c >0$ for all $r \in V_+$. 
This implies $n\in \ovl{V_+}\setminus\{0\}$ and $c \ge 0$.
Hence there is a $b \in \bC$
with $\Im b = c \ge 0$ 
such that $n\cdot \zeta +b = 0$ (while $\Im (n\cdot q +b) > 0$
for all $q \in {\rm T}_+$).
This proves Lemma \ref{hw} for all dimensions.
\fi

\ifundefined{GlaserApp}{}\else
\section{Appendix. Sketch of the proof of Glaser's theorem 1}
\label{glproof}
This section closely follows the original \cite{[G1]} with a few unimportant
alterations, mainly intended for the cases when $U_n$ might not be
simply connected. 
The notations are those of Glaser's theorem 1, and we also denote 
$\wt U_n$ the universal covering space of $U_n$, $\iota_n$ the canonical
projection of $\wt U_n$ onto $U_n$. If $\VV$ is a complex manifold,
$\AA(\VV)$ denotes the set of holomorphic functions on $\VV$.
It is clear that (G.1) $\Rightarrow$ (G'.1). The latter 
implies (G.0), by inserting 
$g_n(z_n) = f(z_n - p_n) \delta(\Im (z_n - p_n))$ in G'.1.
In turn (G.0) implies (G.4) by inserting 
$f_n(h_n) = \sum_\alpha a_\alpha(n) \partial^\alpha \delta(h_n)$ in
(G.0). 

\noindent 1. The first step of the proof is to show that 
(G.4) $\Rightarrow$  (G.3). Assume that (G.4) holds. Let, for each
$n \in [1,\ M]$, $R_n >0$ be such that the closure of the polydisk 
$P_n = \{z_n \in \bC^{N_n}\ :\ |z_{n,j}-p_{n,j}| < R_n \forall j\}$
is contained in $U_n$. For any $\{z_n \in P_n\}_{1\le n\le M}$ and 
every finite sequence $b(n)_\alpha$, 
\begin{equation}
Q_{z}(b,\ b) = Q_{p}(a,\ a),
\end{equation}
\begin{equation}
a(n)_\alpha = \alpha!\,\sum_{\gamma \le \alpha} 
{b_\gamma (z_n-p_n)^{\alpha - \gamma} \over \gamma!\,
(\alpha - \gamma)!}
\end{equation}
Although the sequence $\{a(n)_\alpha\}_{\alpha \in \bN^{N_n}}$ is infinite,
the convergence of the power series for $Q_{z}(b,\ b)$ and a limiting argument
show that $Q_{z}(b,\ b) \ge 0$. Thus the property (G.4) propagates 
everywhere, i.e. (G.3) holds.

\noindent 2. As our next step, we prove that (G.4) implies that (G'.1) holds 
within the same sequence of polydisks $\{P_n\}_{n \in [1,\ M]}$ just used.
We first prove this in the form of condition (G''.1), i.e. in
case $g_n$ is a finite linear combination of
Dirac measures, i.e.
\begin{equation}
g_n(z_n) = \sum_{r = 1}^L c_{n,r} \delta(z_n - t_{n,r}) ,
\end{equation}
with $t_{n,r} \in P_n$. Indeed 
\begin{equation}
\sum_{n,\ m = 0}^M\,
\sum_{r,\ s = 1}^L\, 
c_{n,r} \ovl{c_{m,s}} A_{n\ m}(t_{n,r},\ \ovl{t_{m,s}}) =
Q_p(a,\ a),
\end{equation}
with
\begin{equation}
a(n)_\alpha =
\sum_{r=1}^L c_{n,r} (t_{n,r} - p_n)^\alpha.
\end{equation}
The sequence $\{a(n)_\alpha\}$ is again infinite, but 
we can still conclude that $Q_p(a,\ a) \ge 0$,
i.e. that (G''.1), and hence (G'.1) hold in the sequence of domains 
$\{P_n\}_{n \in [1,\ M]}$.

\noindent 3. To prove that (G'.1) in the 
sequence of polydisks  $\{P_n\}_{n \in [1,\ M]}$  
implies the property (G.2) within the same sequence, we introduce a Hilbert
space $E = E_0 \oplus \ldots E_M$ as follows: $E_0 = \bC$. For each
$n \in [1,\ M]$, $E_n = \AA(U_n) \cap L^2(P_n,\ \lambda_n)$. It is 
well-known (see [B]) that $E_n$ is a closed subspace of 
$L^2(P_n,\ \lambda_n)$, and that the convergence of a sequence 
$\{\psi_\nu \in E_n\}_{\nu \in \bN}$ in the sense of  $E_n$ implies
its uniform convergence on every compact subset of  $P_n$
The operator $A_P$ defined on $E$ by
\begin{equation}
(g,\ A_P\,f) = \sum_{m,\ n=0}^M
\int_{P_n \times P_m} \ovl{g_n(p_n)}\,A_{n\ m}(p_n,\ \bar q_m)\,
f_m(q_m)\, d\lambda_n(p_n)\,d\lambda_m(q_m)
\end{equation}
is Hilbert-Schmidt and positive by virtue of the property (G'.1). The
spectral decomposition of this operator therefore yields the existence
of a sequence 
$\{\vhi_\nu = (\vhi_{\nu,\ 0},\ldots,\ \vhi_{\nu,\ M})\ :\ \nu \in \bN\}$
of eigenvectors of $A_P$ corresponding to non-negative eigenvalues. Hence
there is a sequence $\{f_\nu \in E\}$ such that 
\begin{equation}
A_{n\,m}(p_n,\ \bar q_m) = \sum_{\nu \in \bN}
f_{\nu,\ n}(p_n)\,\ovl{f_{\nu,\ m}(q_m)}
\label{locexp}\end{equation}
holds for all $n,\ m \in [0,\ M]$, uniformly on every compact subset of
$P_n\times P_m$.

\noindent 4.
We now show that for each $n$ and $\nu$, $f_{\nu,\ n}$ extends to a
function holomorphic on $\wt U_n$. Let $z_n \in P_n$ and suppose that
the closure of
a polydisk $P'_n$ with radius $R'$ centered at $z_n$ is contained in
$U_n$. The Taylor coefficients of $A_{n\ n}$ at $z_n$ satisfy
\begin{equation}
{\partial^\alpha\,\bar \partial^\alpha \over \alpha !^2}
A_{n\ n}(z_n,\ \bar z_n) = 
\sum_\nu \left | {\partial^\alpha f_{\nu,\ n}(z_n) \over \alpha !}\right |^2
\ \ \le C\,R'^{2|\alpha|}.
\label{137}\end{equation}
Hence the power series for $f_{\nu,\ n}$ at $(z_n)$ converges in 
$P'_n$. Moreover the expansion (\ref{locexp}) continues to hold in
$P'_n \times P'_m$. To see this it suffices, by Schwartz's inequality,
to prove that $\sum_\nu |f_{\nu,\ n}(w_n)|^2$ converges in $P'_n$.
We fix $n \ge 1$ and temporarily denote 
$g_{\nu,\ \alpha} = \partial^\alpha f_{\nu,\ n}(z_n)/\alpha!$. 
For any $\kappa \in (0,\ 1)$, by Schwartz's inequality,
\begin{equation}
\left |
\sum_\alpha |g_{\nu,\ \alpha}| (\kappa^2 R')^{|\alpha|}
\right |^2 \le
(1-\kappa^2)^{-N_n}\ 
\sum_\alpha |g_{\nu,\ \alpha}|^2 (\kappa^2 R'^2)^{|\alpha|} ,
\end{equation}
and by Eq. (\ref{137}),
\begin{equation}
\sum_\nu \left |
\sum_\alpha |g_{\nu,\ \alpha}| (\kappa^2 R')^{|\alpha|}
\right |^2 \le
C (1-\kappa^2)^{-2N_n}.
\end{equation}
In particular for any $\veps >0$,
it is possible to chose $S$ such that 
\begin{equation}
\sum_{\nu \ge S} \left |
\sum_\alpha |g_{\nu,\ \alpha}| (\kappa^2 R')^{|\alpha|}
\right |^2 \le \veps
\end{equation}
so that, for any $\zeta$ with $|\zeta|< \kappa^2 R'$,
\begin{equation}
\sum_{\nu \ge S} |f_{\nu,\ n}(z_n +\zeta)|^2 \le \veps ,
\end{equation}
\begin{equation}
\sum_\nu |f_{\nu,\ n}(z_n +\zeta)|^2 \le C (1-\kappa^2)^{-2N_n}.
\end{equation}
Therefore there exists a function $\wt f_{\nu,\ n}$ holomorphic
on $\wt U_n$ and a component $\wh P_n$ of $\iota_n^{-1}(P_n)$ such that
$\wt f_{\nu,\ n}$ coincides with $f_{\nu,\ n}\circ \iota_n$ on 
$\wh P_n$. The expansion 
\begin{equation}
\wt A_{n\ m}(\zeta_n,\ \bar \zeta_m) 
\buildrel def \over {=}
A_{n\ m}(\iota_n(\zeta_n),\ \ovl{\iota_m(\zeta_m)}) = 
\sum_\nu \wt f_{\nu,\ n} (\zeta_n)\,
\ovl{\wt f_{\nu,\ m} (\zeta_m)}
\end{equation}
holds uniformly on every compact subset of $\wt U_n \times \wt U_m$.
Therefore the sequence $\{\wt A_{n\ m}\}$ posesses the property 
(G''.1) in $\{\wt U_n\}$. It follows that the $\{A_{n\ m}\}$ possess
the property (G''.1) in $\{U_n\}$, since the points $t_{n,\ r}$ can
be lifted in an arbitrary way to points in $\wt U_n$.

\noindent 5.
For each $n \in [1,\ M]$,
we now define a measure $\rho_n$ on $U_n$ 
as follows. Let first 
$d\mu_n(p_n) = \e^{-\vhi_n(p_n)}\,d\lambda_n(p_n)$ where the
smooth real function $\vhi_n$ is chosen such that 
$\int_{U_n}  d\mu_n(p_n) = 1$. We then define 
$d\rho_n(p_n) = (1+A_{n\ n}(p_n,\, \bar p_n))^{-1}\,d\mu_n(p_n)$.
Let $F_0 = \bC$, and, for each $n \in [1,\ M]$, let
$F_n = \AA(U_n)\cap L^2(U_n,\ \rho_n)$. Note, e.g. that for any
fixed $q_n \in U_n$, the function 
$p_n \mapsto A_{n\ n}(p_n,\, \bar q_n)$ belongs to $F_n$.
Let $F = F_0 \oplus \ldots \oplus F_M$.
The operator ${\bf A}$ defined on $F$ by
\begin{equation}
(g,\ {\bf A}\,f) = \sum_{m,\ n=0}^M
\int_{U_n \times U_m} \ovl{g_n(p_n)}\,A_{n\ m}(p_n,\ \bar q_m)\,
f_m(q_m)\, d\rho_n(p_n)\,d\rho_m(q_m)
\end{equation}
is Hilbert-Schmidt and positive, and we again conclude that there
exists a sequence $\{h_\nu \in F\ :\ \nu \in \bN\}$ such that
\begin{equation}
A_{n\ m}(p_n,\ \bar q_m) = \sum_{\nu \in \bN}
h_{\nu,\ n}(p_n)\,\ovl{h_{\nu,\ m}(q_n)}
\end{equation}
holds uniformly on every compact subset of $U_n\times U_m$ as well as
in the sense of $F_n \otimes F_m$.

This concludes the proof of Glaser's theorem 1.

\noindent{\bf Remarks}\HB
1. The extension to the case when the $U_n$ are complex manifolds 
which are separable at infinity is straightforward.\HB
2. The requirement that $U$ be simply connected in Glaser's theorem 2  
is necessary
as the following example shows. Let $U = \bC \setminus \{0\}$ and
$A(p,\ \bar p) = |p|$. This satisfies the assumptions of Glaser's theorem 2
since $A(p,\ \bar p) = \sqrt{p}\,\sqrt{\bar p}$. But there cannot be a 
sequence $f_\nu$ of functions holomorphic on $U$ such that
$A(p,\ \bar p) = \sum_\nu |f_\nu(p)|^2$, since $|f_\nu(p)| \le \sqrt{|p|}$
implies that $f_\nu$ is analytic at 0, hence entire and necessarily 0.
\fi

\end{document}